\documentclass[sn-basic]{sn-jnl}% Math and Physical Sciences Reference Style

\jyear{2024}%

\usepackage{sidecap}
\theoremstyle{thmstyleone}%

\theoremstyle{thmstyletwo}%

\theoremstyle{thmstylethree}%

\newcommand{\tr}{^{\sf T}}
\begin{document}

\title[chiPower transformation for CoDA]{The chiPower transformation: a valid alternative to logratio transformations in compositional data analysis}

\author*[1,2]{\fnm{Michael} \sur{Greenacre}}\email{michael.greenacre@gmail.com}


\affil*[1]{\orgdiv{Department of Economics and Business, and Barcelona School of Management}, \orgname{Universitat Pompeu Fabra}, \orgaddress{\street{Ramon Trias Fargas, 25--27}, \city{Barcelona}, \postcode{08272}, \country{Spain}}}

%\author*[1,2]{\fnm{\ } \sur{\ }}\email{\ }


%\affil*[1]{\orgdiv{\ \qquad\qquad\qquad\ }, \orgname{\ \qquad\qquad\qquad\ }, \orgaddress{\street{\ \qquad\qquad\qquad\ }, \city{\ \qquad\qquad\qquad\ }, \postcode{\ \qquad\qquad\qquad\ }, \country{\ \qquad\qquad\qquad\ }}}

%\affil[2]{\orgdiv{Department}, \orgname{Organization}, \orgaddress{\street{Street}, %\city{City}, \postcode{10587}, \state{State}, \country{Country}}}


\abstract{The approach to analysing compositional data has been dominated by the use of logratio transformations, to ensure exact subcompositional coherence and, in some situations, exact isometry as well.
A problem with this approach is that data zeros, found in most applications, have to be replaced to allow the logarithmic transformation.
An alternative new approach, called the `chiPower' transformation, which allows data zeros, is to combine the standardization inherent in the chi-square distance in correspondence analysis, with the essential elements of the Box-Cox power transformation. 
The chiPower transformation is justified because it} defines between-sample distances that tend to logratio distances for strictly positive data as the power parameter tends to zero, and are then equivalent to transforming to logratios.
For data with zeros, a value of the power can be identified that brings the chiPower transformation as close as possible to a logratio transformation, without having to substitute the zeros.
Especially in the area of high-dimensional data, this alternative approach can present such a high level of coherence and isometry as to be a valid approach to the analysis of compositional data.
Furthermore, in a supervised learning context, if the compositional variables serve as predictors of a response in a modelling framework, for example generalized linear models, then the power can be used as a tuning parameter in optimizing the accuracy of prediction through cross-validation.
The chiPower-transformed variables have a straightforward interpretation, since they are each identified with single compositional parts, not ratios.


\maketitle

\section{Introduction}\label{sec1}

Compositional data are non-negative data carrying relative, rather than absolute, information. Often these data have a constant-sum constraint on each sample's set of values, for example, proportions summing to 1 or percentages summing to 100\%.
Such data are found in many fields, notably biochemistry, geochemistry, ecology, linguistics, as well as all the  ``omics" fields of genomics, microbiomics, transcriptomics, metabolomics, etc.. 
In most cases, such data are originally observed as counts, abundances or intensities, where the totals in the samples, usually the row totals of the original data matrix, are irrelevant.
Consequently, the sample values can be divided by their respective totals to give vectors, called compositions, with sums equal to 1.
This operation of dividing by the total is called closing, sometimes referred to as normalization. 

It has long been recognized that such data need special statistical treatment, since the values in the compositions would change if some compositional parts were excluded and the data re-closed with respect to their new totals, giving so-called subcompositions.
In reality, in almost all applications the observed compositions are themselves subcompositions of a larger set of potentially observable parts, with proportional values that would change if an extended set of parts were observed. 
For example, in geochemistry, some studies use only major oxide elements, others treat trace elements, while others treat the full lithogeochemical spectrum of major, minor, trace and rare elements. Thus, in this last case, the compositional proportions of the major oxides would be different than those when the major oxides were studied alone. 
Similarly, in the study of fatty acid compositions in  biochemistry, the set of fatty acids identified and analysed in any study is always a subcomposition of a much larger set, not only due to the focus of the research but also on the sophistication of the measuring instruments (e.g., gas chromatographs). 
The same is true for microbiome studies, for example, where the set of bacteria is never the full set of possibilities.  
One of the few contexts where a full composition is observed is in daily time use in behavourial studies, where all activities are recorded over a full 24-hour period -- here the time budget is compositionally  complete since no more time can be added to a day.

To deal with this dependency of compositional data on the particular set of parts that are included, the use of ratios of parts as the basis for statistical analysis was proposed by John Aitchison \citep{Aitchison:82, Aitchison:86}, who laid the foundation for a field of statistics often referred to as compositional data analysis, or CoDA.
Ratios are invariant with respect to deleting parts from or adding parts to a composition, and are thus described as being subcompositionally coherent (simply referred to here as coherent), whereas any analysis of the original compositional data is incoherent.
But ratios are awkward to handle statistically -- their distributions are generally skewed and there is an asymmetry between the numerator and the denominator so that, for example, the variance of $A/B$ is not equal to the variance of $B/A$.
The logarithmic transformation reduces the skewness, the variance of $\log(A/B)$ equals the variance of $\log(B/A)$, and either $\log(A/B)$ or $\log(B/A)$ can be used in linear modelling, since they are just a change of sign. 
 Because of the logarithmic transform, additive changes in logratios are thus multiplicative changes in the ratios, as in logistic regression, for example, where a logratio, the log-odds, is modelled as an additive model of explanatory variables, and additive effects back-transform to multiplicative effects on the odds.
 
Hence, logarithms of ratios, called logratios, have become the preferred transformation for those following the tradition of Aitchison, and once this transformation is made, regular statistical methods applicable to interval-scale data can continue as before.
 This approach is exemplified by \cite{Grunskyetal:24}, who present a workflow called GeoCoDA for using the logratio transformation in both unsupervised and supervised learning in geochemistry.
For an in-depth review and reappraisal of Aitchison's ideas and legacy in the 40 years since his 1982 JRSS discussion paper \citep{Aitchison:82}, see \cite{GreenacreEtAl:23}.

Coherence is the main advantage of the logratio approach, but its main disadvantage is the problem of data zeros, as well as the interpretation of results involving logratios.
Data zeros need to be replaced before logratios can be computed, and there have been many proposals to do so, for example \cite{Palarea:15, Lubbe21}.
It may be that alternative transformations, with simpler interpretations and natural handling of data zeros, are close enough to this ideal property of coherence for all practical purposes.
To quantify this ``closeness" to coherence, a possible measure of incoherence has already been proposed by \cite{Greenacre:11a}, using a concept from multidimensional scaling called stress.  
In the present paper, an alternative measure will be used based on the Procrustes correlation, a by-product of Procrustes analysis (see Appendix 2), since this will unify the treatment of coherence and another concept called isometry.

Whereas coherence is a property of the compositional parts, isometry is a property of the samples. 
If the logratio approach is taken as a favourable reference for CoDA, then the sample structure using an alternative transformation can be checked against the sample structure using the logratio transformation. 
Here the Procrustes correlation will again be used to measure closeness to isometry, by which is meant closeness to the logratio sample structure. 
This idea of using Procrustes analysis, inspired by \cite{Krzanowski:87}, has already been used for logratio variable selection by \cite{Greenacre:19}. 
Such diagnostic measures of similarity between part structures (coherence) and between sample structures (isometry) allow practitioners to judge whether simpler alternative transformations are close enough to coherence and isometry to allow valid statistical analysis.
As mentioned before, the benefit of these alternative transformations will be that they are easier to interpret and also cope naturally with zeros in the data without need for replacement or imputation.

The objective of this paper is to demonstrate how the intrinsic standardization in correspondence analysis \citep{Benzecri:73, Greenacre:84, Greenacre:16a}, followed by a Box-Cox power transformation \citep{Box:64}, can be successfully used as an alternative to logratio transformations.
This alternative is underpinned by the fact that correspondence analysis's chi-square distances computed on Box-Cox transformed compositions tend to logratio distances as the power parameter tends to zero \citep{Greenacre:09, Greenacre:10a}.
This close theoretical connection holds for strictly positive data, and clearly not for data that include zeros.
However, in the presence of zeros, it turns out that a power transformation can be identified that is optimal in approximating  logratio distances (i.e., as close to isometry as possible), and the validity of the resulting transformation can be additionally checked using the measure of coherence, for comparison of various subcompositions to the full composition.
Because the proposed transformation combines the ideas of chi-square standardization (i.e., division of the part values by the square roots of their respective mean values) and power transformation, the new transformation is termed the \textit{chiPower} transformation, to be defined explicitly in Subsection 2.3 below.

Moreover, if the compositional variables serve as independent variables in a supervised learning context, then the value of the power can be used as a tuning parameter to optimize prediction of the response variable.
In this particular situation isometry is no longer important, but coherence is still an issue and will need to be investigated in each case.  

To illustrate this alternative approach, a ``wide" compositional data set is first considered with almost 4000 compositional parts (microbial genes) \citep{Martinez:22, GreenacreMartinezBlasco:21}. 
This is a typical data set in the burgeoning field of ``omics" research: genomics, microbiomics, metabolomics, proteomics, etc.
A second data matrix with much fewer parts but many more samples, i.e. a ``narrow" but ``long" data set, is considered where there is a categorical response to be predicted from the compositional variables.
In both applications the issue of data zeros is considered.

%%%%%%%%%%%%%%%%%%%%%%%%%%%%%%%%%%%%%%%%%%
\section{Material and Methods}\label{sec2}

\subsection{Data sets ``Rabbits" and ``Crohn"}
To demonstrate the suitability of the chiPower approach proposed here, two data sets are considered:
\begin{enumerate}
  \setlength\itemsep{1em}
  \item Data set ``Rabbits", used by \cite{GreenacreMartinezBlasco:21}: a ``wide" data set of counts of $J=3937$ microbial genes observed on a sample of $I=89$ rabbits.  The advantage of this data set is that it has no zero values, so the logratio transformations are valid on all the data.  By simulating a large percentage of small counts to be zeros, the behaviour of the chiPower transformation, which can handle zero values without any problem, can be studied in comparison with the original logratio-transformed data.
  
  \item Data set ``Crohn", used by \cite{Calle:11} and available in the \textsf{R} package \texttt{coda4microbiome} \citep{Calle:23}. This is a ``narrow" matrix of counts of bacterial species aggregated into $J=48$ genera on $I=975$ human samples.  In addition, each sample has been classified as having the digestive ailment called Crohn's disease (662 samples) or not (313 samples). A curiosity of this data set is that it has been published in two versions: first, the original one with many data zeros (totalling 13474, i.e., 28.8\% of the data set), in the original \texttt{selbal} \textsf{R} package -- this version was analysed by \cite{Rivera:18} (see Supplementary Material Section S1), who explicitly state that the ``replacement of zeros by positive numbers is performed under the assumption that the observed zeros represent rounded zeros"; and second, a modified version published in the  \texttt{coda4microbiome} package, with the same data set name \texttt{Crohn}, where the value of 1 has been added to all the counts, no doubt to avoid the zero problem when computing logratios. As of the date of writing, no warning or explanation in the \texttt{coda4microbiome} package is given that the data set has been changed in this way, where $975\times 48 = 46800$ counts have effectively been added to the data set. 
  
  Nevertheless, the advantages of considering both versions of these data are two-fold. First, thanks to the large number of samples, a machine-learning approach can be applied to both versions for predicting the disease, where cross-validation can be implemented to estimate prediction accuracy; and second, the original data set, without zero replacement, can be used to show how well the chiPower approach, applied to the original data with zeros, compares to the logratio approach applied to the modified data set without zeros.
  Since other papers may have used the original \texttt{Crohn} data handling the zeros in different ways, the issue of the effect of these zero replacement strategies on the data variance is dealt with in Supplementary Material Section S1.
  The two versions of the data will be referred to as ``the original Crohn data, with zeros" and ``the modified Crohn data, without zeros".
\end{enumerate}

\subsection{Logratio transformations}
Because the new chiPower transformation will be compared with the logratio approach, a short summary of the most relevant logratio transformations is given here \citep{Aitchison:86}.
Suppose $\bf X$ is an $I\times J$ samples-by-parts (closed) compositional data matrix, and $[x_1 \  x_2 \ \cdots \  x_J]$ is a general row of $\bf X$, that is, a $J$-part composition, where $\sum_{j=1}^J x_j = 1$.
A specific row, for example the $i$-th row of $\bf X$, is denoted  $[x_{i1} \  x_{i2} \ \cdots \  x_{iJ}]$.

The basic logratio transformation is the pairwise logratio transformation, denoted by LR, of two parts $j$ and $j^\prime$
\begin{equation}
{\rm LR}(j,j^\prime) = \log(x_{j}/x_{j^\prime})    
\label{LR}
\end{equation}
There are $J(J-1)/2$ unique LRs, but only $J-1$ linearly independent ones are needed to generate all the others by linear combinations \citep{Greenacre:18}.
Thus, for $I$ compositional samples, the $I\times J(J-1)/2$ matrix of LRs has rank $J-1$.

A special case of LRs are the additive logratios (ALRs), where the denominator part (also called the reference part, ref) is fixed.
\begin{equation}
{\rm ALR}(j\vert\textrm{ref}) = \log(x_{j}/x_{\textrm{ref}}), \quad j=1,\ldots, J, \ j\neq \textrm{ref}    
\end{equation}
There are $J$ choices for the reference part, each of which gives $J-1$ ALRs.
Any $I\times (J-1)$ data matrix of ALRs has rank $J-1$, and the choice of the reference part is determined by domain knowledge, or based on a statistical criterion such as the one that gives a transformed matrix closest to being isometric, or the one with lowest variance of its log-transform \citep{GreenacreMartinezBlasco:21}.
In the latter case, if the variance of $\log(x_\textrm{ref})$ is low, i.e. $\log(x_\textrm{ref})$ is nearly constant, then the ALR  $\log(x_{j}/x_{\textrm{ref}}) = \log(x_{j}) - \log(x_{\textrm{ref}})$ is an approximate constant shift from the $\log(x_{j})$ values themselves, in which case the ALRs can be more easily interpreted as close to the logarithm of the numerator parts.

The centered logratio (CLR) transformation is the log-transform of each part divided by the geometric mean of all the parts:
\begin{equation*}
{\rm CLR}(j) = \log(x_{j}/(x_1 x_2 \cdots x_J)^{1/J}), \quad j=1,\ldots, J 
\end{equation*}
There is only one set of $J$ CLRs and the $j$-th one is the average of all the LRs $\log(x_j/x_{j^\prime})$, for $j^\prime = 1, 2, \ldots, J$, one of which, $\log(x_j/x_{j})$, is zero. 
The $I\times J$ data matrix of CLRs also has rank $J-1$, due to a linear relationship amongst them. They are generally not used as variables representing the individual parts, although it is tempting to do so, but rather as representing all the LRs by their differences: ${\rm LR}(j,j^\prime) = {\rm CLR}(j) - {\rm CLR}(j^\prime)$. 
For example, to construct the sample logratio geometry, by which is meant the Euclidean distance structure of the samples with respect to all LRs, it is not necessary to work with the $I\times J(J-1)/2$ matrix of all LRs, but just with the $I\times J$ matrix of CLRs \citep{AitchisonGreenacre:02}. The logratio distances between samples using the CLRs are identical to those using all the LRs \citep{Greenacre:18, Greenacre:21}.

Transforming by logratios takes the compositions inside the simplex out into real vector space, where regular interval-scale statistical analysis, both univariate, bivariate and multivariate can be performed.
%In particular, distances between compositions can be computed, using Euclidean distances on their $J(J-1)/2$ LRs, or equivalently on their $J$ CLRs, which is more efficient.
The problem, however, is with data zeros, which need replacement before such transformations can be made.

\subsection{The chiPower transformation: chi-square standard- ization, with preliminary power transformation}
In correspondence analysis (CA), usually applied to a matrix of counts, the rows are first divided out by their totals to get so-called row profiles, synonymous with compositions -- see, for example, \cite{Greenacre:16a}. 
In CoDA terminology, CA automatically closes the rows, and -- if the analysis is considered column-wise -- it symmetrically closes the columns to get column profiles. 
In a closed compositional data matrix, the compositions in the rows are already profiles, so closing in CA does not change them.
The row profiles in CA are weighted proportionally to the original marginal row totals, but in the case of compositions these marginal sums are all equal, so there is uniform weighting on the rows.
Finally, distances between profiles in CA are chi-square distances, which are Euclidean distances after standardizing each compositional value $x_j$ by dividing by the square root of its expected value, the column (part) mean $\bar{x}_j$ (see, for example, \cite{GreenacrePrimicerio:13}, chapter 4): i.e., $x_j / \sqrt{\bar{x}_j}$ -- this is called the chi-square standardization. 
In the chiPower transformation, the $x_j$ will be raised to power $\lambda$ and closed, again giving compositions, and then divided by their respective columns means. 
%Notice that the variances of chi-square standardized parts are equal to 1 for (scaled) Poisson distributed data .

%From the row profiles' ``point of view", CA of the compositional data matrix is equivalent to a principal component analysis (PCA) of the profile matrix standardized in the above way.
%In a low-dimensional PCA biplot, where rows (samples) are in principal coordinates (thus, approximating the chi-square distances) and the columns, which have equal normalization, are in contribution coordinates \citep{Greenacre:13} (thus, distinguishing which parts have a higher or lower contribution to the biplot).

For the present purpose, the Box-Cox power transformation is defined for positive $x$ as:
\begin{equation}
     f(x \, \vert \, \lambda) =
    \begin{cases}
      \frac{1}{\lambda}\left(x^\lambda -1\right) & \text{if\ } \lambda > 0 \\
      \log(x) & \text{if\ } \lambda = 0
    \end{cases}
  \label{BoxCox}    
\end{equation}
%where $f(x \, \vert \, \lambda)$ tends to $\log(x)$ as $\lambda$ tends to 0 
(negative values of $\lambda$ are not considered, and only values $0 < \lambda \leq 1$ are of present interest).
Whereas the limit result implicit in (\ref{BoxCox}), that is, $f(x \, \vert \, \lambda) \to  \log(x) \rm{\ as\ } \lambda \to 0$, is only valid for $x > 0$, the power transformation itself for $\lambda > 0$ is valid for nonnegative $x$, i.e. $x \geq 0$, which is the way it will be used in the present approach. 
The subtraction $-\frac{1}{\lambda}$ in (3) is not usually required: 
for example, for computing the chi-square distances between samples, the differences between row profiles (i.e., compositions) that have been transformed according to (\ref{BoxCox}) will eliminate the constant shift term $-\frac{1}{\lambda}$ in the distance formula.
Closing of the samples (rows) in the power-transformed data matrix turns it into a compositional data matrix, 
in line with the standard operation in CoDA called ``powering", defined by \citep{Aitchison:86} as the closure of a power-transformed compositional data matrix.
The scale factor $\frac{1}{\lambda}$, however, is important to retain, especially in unsupervised learning approaches, otherwise the variance in the transformed (positive) data reduces as $\lambda$ decreases. 
 As shown in Appendix 1, if one wants the chiPower transformation to converge directly to the CLR transform, then a scale factor of $\sqrt{J}$ needs to be introduced and the $-1$ of the Box-Cox transform needs to be retained. 

%The proposal here is thus to apply the power transformation $\left(x^\lambda\right)$ to each value in the compositional data matrix, and then standardize each part by the square root of the transformed part means.
%
%Introducing the rescaling factor $\frac{1}{\lambda}$ is necessary, for the reason given above.

The chiPower transformation, which thus combines the chi-square standardization with the Box-Cox style of powering, and converges to the CLR transform, is defined algorithmically in the following steps, where the determination of the power $\lambda$ will be dealt with after the definition.

\medskip

\leftline{\bf The chiPower transformation:}
\begin{enumerate}
    \item For a given $\lambda$, power transform the compositional data matrix $\bf X$ to obtain ${\bf X}{\scriptstyle [\lambda]} = \big[x_{ij}^\lambda\big]$, where $0 < \lambda \leq 1$ (so the possibility of no power transformation is included, when $\lambda=1$).
    \item Close the rows of $\bf X{\scriptstyle [\lambda]}$ to obtain another matrix of compositions, $\bf Y{\scriptstyle [\lambda]}$ 
    \item Compute the vector of column means $\bar{\bf y}{\scriptstyle [\lambda]} = \big[ \bar{y}{\scriptstyle [\lambda]}_1 \  \bar{y}{\scriptstyle [\lambda]}_2  \cdots  \bar{y}{\scriptstyle [\lambda]}_J \big]$ of ${\bf Y}{\scriptstyle [\lambda]}$.
    \item Divide the columns of the closed ${\bf Y}{\scriptstyle [\lambda]}$ by the square roots of the respective column means (i.e., the chi-square standardization) and apply the Box-Cox style of transformation as follows: 
    \begin{equation}
        z_{ij}{\scriptstyle [\lambda]} = \frac{1}{\lambda} \big( \sqrt{J}\frac{y_{ij}{\scriptstyle [\lambda]}}{\sqrt{\bar{y}{\scriptstyle [\lambda]}_j}} - 1\big)
        \label{chiPower}
    \end{equation} 
    The inclusion of the scale factor $\sqrt{J}$ is related to the convergence to the CLR transformation and is shown in Appendix 1.
    \item ${\bf Z}{\scriptstyle [\lambda]} = \big[z_{ij}{\scriptstyle [\lambda]}\big]$ is the chiPower-transformed data matrix with power $\lambda$.
    Euclidean distances between the rows of ${\bf Z}{\scriptstyle [\lambda]}$ are called chiPower distances between the rows of $\bf X$, which for $\lambda=1$ are the chi-square distances in a regular CA context. The set of all Euclidean distances between rows of ${\bf Z}{\scriptstyle [\lambda]}$, i.e. the Euclidean geometry of chiPower-transformed data, defines the chiPower geometry of the original matrix $\bf X$.
    \end{enumerate}
\vspace{-0.2cm}
 As shown in Appendix 1, the chiPower transformation converges in the limit, as $\lambda$ tends to 0, to the CLRs that have been negatively shifted by the column means of ${\bf Z}{\scriptstyle [\lambda]}$. This can be corrected to give actual CLRs in the limit, if required, by simply adding the column means of ${\bf Z}{\scriptstyle [\lambda]}$. This is done by default in the \textsf{R} function \texttt{chiPower}.
\noindent

The way the power $\lambda$ is chosen will depend on the statistical learning objective. 
In unsupervised learning, the power can be chosen to make the chiPower geometry of the samples be as close as possible to their logratio geometry (see Section 2.4). This means that methods such as PCA and clustering of the samples can be validly performed on the chiPowered data, as a substitute for logratio-transformed data.
In supervised learning where the compositions serve as predictors of a response, $\lambda$ will be chosen to optimize model fit or predictivity, and if the sample is of sufficient size, the power can be chosen by cross-validation.
In this case, not all the above steps are necessary -- for example, steps 3 and 4 only change the scales of the predictors linearly and this does not affect their roles in modelling.
In supervised learning where the compositions serve as responses, however, not only would closeness to logratio geometry be important, but also the predictability of the compositions by the explanatory variables -- in this case a compromise would perhaps be desirable in choosing $\lambda$ between these competing objectives.

The idea to apply the Box-Cox style of power transformation to compositional data is not new -- see \citep{Aitchison:86, Rayens:91, Tsagris:16}.
\cite{Greenacre:10a} showed the connection of Box-Cox transforming prior to performing CA with logratio analysis (LRA, i.e. the PCA of CLR-transformed data). 
In the present work, however, we use this idea in a much wider context of analysing compositional data, both unsupervised and supervised.
 A recent paper by \cite{Erb:23} also looks at estimating the power parameter of power-transformed compositions, considering this as a shrinkage problem, even proposing to estimate a different power for each sample. Estimating a different power for each compositional part is a further possibility, since each part has a different level of skewness.

Furthermore, Section S3 of the Supplementary Material shows how CA applied to a closed power-transformed data matrix, where the samples (rows) are equally weighted, reduces to a PCA of the chiPower-transformed data.
The only difference between the two analyses is the treatment of the scalar factor $\frac{1}{\lambda}$, which is eliminated in CA and so has to be re-introduced into the final CA results.

\subsection{Measuring closeness to isometry}
Isometric means ``the same metric", that is the same distance structure in multivariate space.
In the present context, the term applies to the comparison with the reference sample geometry based on logratio distances, which are the Euclidean distances computed on the CLRs -- see Section 2.2. Notice that the specific definition of logratio distance by \cite{Greenacre:21} allocates weights to both the samples and the  compositional parts, where equal weights are used here for both rows and columns.

Consider the logratio distances between all the samples as the reference, and the distances between the same samples based on any other transformation of the compositional data, which in the present case is selected from the family of chiPower transformations.
The closeness of the sample geometry of chiPower-transformed data to the sample logratio geometry can be measured by the Procrustes correlation between the respective sample configurations (Appendix 2 explains how this correlation is obtained).
A convenient way to do this is to apply PCA to the CLR-transformed data and chiPower-transformed data respectively, obtain the complete set of principal coordinates of each, and then fit these two coordinate matrices together by Procrustes analysis.
If the Procrustes correlation is close to 1, this means that the transformation is close to being isometric (always with respect to the logratio geometry, taken as the reference.)

Isometry is important in unsupervised learning, when the structure of the compositional data is being explored by methods such as dimension reduction and clustering, in which case it will be favourable to be close to the logratio geometry, which is known to be coherent. 
It can also be important in supervised learning when the compositions serve as responses to additional explanatory variables, since it is the complete compositional structure that is being modelled. This case is not considered in this paper, but see \cite{Yoo:22} for an application.

\subsection{Measuring closeness to coherence}
Whereas isometry is a property of the samples, coherence is a property of the compositional parts, usually the columns of the data matrix. 
Using LRs and their special case, the ALRs, is a perfectly coherent strategy: for example, LRs involving pairs of parts A, B and C are not affected if additional parts D and E are added to the composition.  

There is nevertheless a relationship between the two concepts of coherence (of the parts) and isometry (of the samples).
In Appendix 1, explicit convergence of the chiPower transformatio to the CLR transformation is shown.
It follows that, since the logratio transformation is perfectly coherent, a transformation such as the chiPower is converging to isometry and coherence at the same time, as the power of the transformation tends to zero.

Notwithstanding this relationship, it is still useful to quantify the level of coherence in a particular application by comparing results for parts in subcompositions and the same parts in the ``full" compositions of the given data. In each case the parts have been transformed in the same way (in this case, using the same chiPower transformation) but computed on different compositions due to the closing operation.
It is also useful to see how the lack of coherence (i.e., incoherence) is affected by the size of the subcompositions, since the subcompositional values will change more due to closing when there are less parts in the subcomposition than in larger subcompositions. 
The type of results to compare depends on the research problem,  because coherence has a different meaning if the statistical analysis is unsupervised or supervised.

In CoDA there is the symmetric concept of the logratio geometry of the parts: logratios can be computed for each part pairwise across the samples (i.e., $I(I-1)/2$ logratios), and their structure is related in the same way to that of the CLRs of the parts \citep{AitchisonGreenacre:02, Greenacre:21}.
There is more than one way to quantify the geometry of the parts in the chiPower approach.
One way is to simply transpose the data matrix and apply the chiPower transformation as before, in other words chiPower the columns (parts). 
%This approach will be illustrated in the Supplementary Material, Section SX.
Another way, which is adopted here, is to use the geometry of the column principal coordinates in the PCA of the chiPowered data.
This defines a distance geometry on the parts which is equivalent to the covariance structure of the transformed parts \citep{GreenacreEtAlPCA:22}.
For unsupervised learning, this logratio geometry of the transformed compositional parts in many different random subcompositions will be compared to the logratio geometry of the same parts, transformed in the same way, in the full compositional data matrix, again using the Procrustes correlation.
So this is  a similar measure of isometry as between the sample geometries, but between the same parts in the subcomposition and the composition. 
One could say the coherence check is being made by measuring the isometry between the parts.

The algorithm for assessing the coherence can be summarized in the following steps.
\begin{enumerate}
    \item Transform the compositional data matrix {\bf X} using chiPower, for the power $\lambda$ of interest, resulting in ${\bf Z}{\scriptstyle [\lambda]}$
    \item Perform the PCA of ${\bf Z}{\scriptstyle [\lambda]}$ using the SVD $I^{-\frac{1}{2}}{\bf Z}{\scriptstyle [\lambda]} = {\bf U D}_\phi {\bf V}\tr$ (see Supplementary Material Section S2).
    \item The part geometry of all the parts is defined by the coordinates ${\bf G} = {\bf V D}_\phi$.
    \item For any subcomposition ${\bf X}_\textrm{s}$, perform the same chiPower transform to obtain ${\bf Z}_\textrm{s}{\scriptstyle [\lambda]}$.
    \item Perform the PCA on ${\bf Z}_\textrm{s}{\scriptstyle [\lambda]}$ (steps 2. and 3.) and define the geometry of the subcompositional parts from the results of this PCA in the same way as before, i.e., coordinates ${\bf G_\textrm{s}}$.
    \item Compute the Procrustes correlation between ${\bf G_\textrm{s}}$ and the subset of rows of $\bf G$ corresponding to the same subset of parts chosen for the subcomposition. 
\end{enumerate}

\noindent
The above is repeated for many subcompositions of different sizes. 

The previous approach by \cite{Greenacre:09} to measure incoherence used a stress measure common in multidimensional scaling \citep{BorgGroenen:10}, applied to the distances between parts. 
This approach used a ``worst-case scenario" of two-part subcompositions, which might be acceptable for small compositions but is too extreme and unrealistic for larger ones that are generally the case in practice.  Here it is preferred to use a range of subcompositions in the range of 10--90\% of the total number of parts, so that the lack of coherence can also be assessed for subcompositions of different sizes. 

For supervised learning when compositions serve as predictors, this approach of comparing geometries of subsets of parts is no longer important, and coherence would rather be assessed by seeing how the model parameter estimates vary for the subcompositional parts compared to their compositional counterparts, all with the same chiPower transformation.

There are clearly very many possibilities to choose subsets of parts in order to create subcompositions and check for incoherence.
Random subsets of parts can be selected, or it may be that subcompositions in particular applied contexts tend to include the more frequent parts more often than the less frequent ones.
For example, in microbiome research, the more frequent bacteria would always be present across different studies, whereas they would vary in the rarer bacteria that they include.
Similarly, in studies of fatty acid compositions, it is again the rarer fatty acids that might not appear in some studies, depending on the sophistication of the laboratory equipment used in the data collection.

%{\color{red} Notice that CLRs as single variables are not coherent because they involve the full set of parts, so the CLRs on compositions are not compatible with those on any subcompositions. 
%Nevertheless, it is still clarifying to check the level of incoherence of CLRs in the same way, and an example of this is given in the Supplementary Material, Section SX.}

%--------------------------------------------
\subsection{The problem of data zeros}
With the chiPower transformation and measures of closeness to isometry and coherence in place, attention is now turned to compositional data with zeros.
The problem of zeros has been called the ``Achilles heel" of compositional data analysis \citep{Greenacre:21}, since data have to be strictly positive to be able to compute logratios.
Because zeros are usually present in compositional data, and often in large quantities, a number of zero replacement strategies have been developed --- see \cite{Lubbe21} for a review.
The presence of many zeros can cause problems in the analysis \citep{teBeest:21}.

As \cite{Lundborg:23} state: ``we believe that it is generally preferable
to modify the statistical procedure to fit the data rather than vice versa".
Using the chiPower transformation provides an approach to avoid zero replacement, but as the power decreases, an incompatibility with logratios will develop because the transformation of the original zeros leads to very large negative numbers as they approach minus infinity, with a resultant degradation of the metric properties of the data.
In the present approach, for data with zeros, the power of the chiPower transformation will be identified that leads to the transformed data having maximum isometry with the sample logratio geometry. However, zeros will have to be replaced to enable computations of the CLRs, which define the logratio geometry, so there is a slight disparity in the comparison between the chiPower-transformed data that have zeros and the logratio-transformed data that have zeros replaced.
See Supplementary Material Section S1 for further discussion of this issue.

\section{Results} \label{sec3}
\subsection{Unsupervised learning: strictly positive compositions}
The compositional data set ``Rabbits" (89 samples, 3937 genes) has strictly positive values, which
is rather atypical, but it is useful here to illustrate the good properties of the chiPower transformation.
The next subsection treats the case with data zeros.

Logratio analysis (LRA) is first performed on the data and the configuration of the 89 samples established in 88-dimensional multivariate space, one less than the number of samples for this wide data set. This is PCA applied to the CLRs.
Then PCA is performed on the chiPower-transformed data, with powers $\lambda$ descending from 1 in small steps to almost 0, where ``almost" is $\lambda=0.0001$. These analyses are effectively all CAs on closed power-transformed data, as explained in Supplementary Material Section S3. 

Figure \ref{Procr_Unsuper}A shows a plot of the Procrustes correlations between the logratio geometry of the 89 samples and corresponding chiPower-transformed geometry, showing the convergence to 1 as $\lambda$ tends to 0. 
In each case along the curve the 88-dimensional logratio geometry is compared to the 88-dimensional chiPower geometry.
Values indicated are for square root, fourth root and ten thousandth root ($\lambda=0.0001$) transformations.

Figure \ref{Procr_Unsuper}B plots the $89\times88/2 = 3916$ logratio distances between pairs of sample points in the full 88-dimensional space against the corresponding distances for the $\lambda=0.0001$ case, where the almost exact isometry is further shown.

\begin{figure}[h]
\begin{center}
\includegraphics[width=12cm]{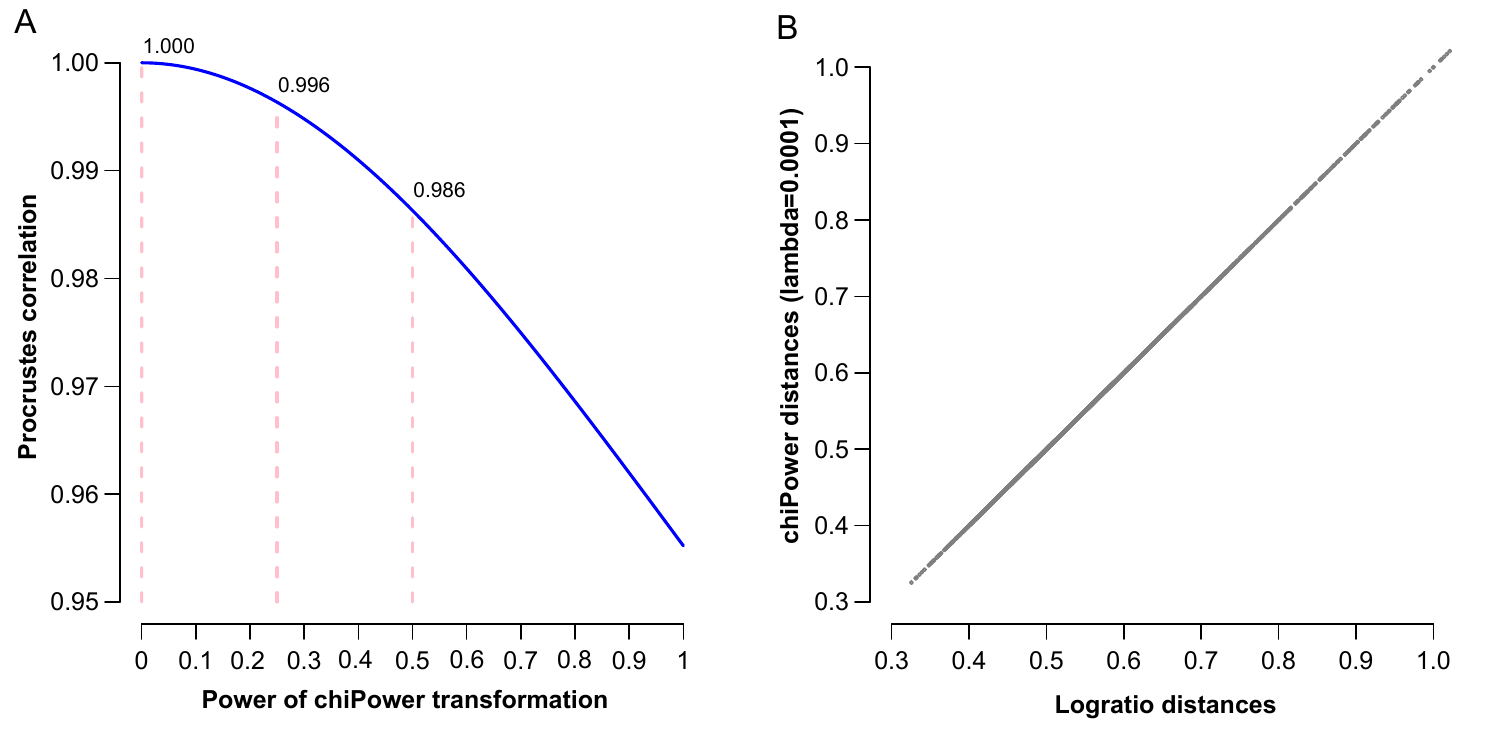}
\caption{A. The Procrustes correlations, measuring proximity to isometry between the exact logratio geometry and the geometry of chiPower-transformed data with different powers, showing the convergence close to 0. B. For the power equal to 0.0001, the chiPower distances are practically identical to the logratio distances. In the limit as the power tends to 0, they are identical.}
\label{Procr_Unsuper}
\end{center}
\end{figure}

\begin{figure}[ht]
\begin{center}
\includegraphics[width=11.8cm]{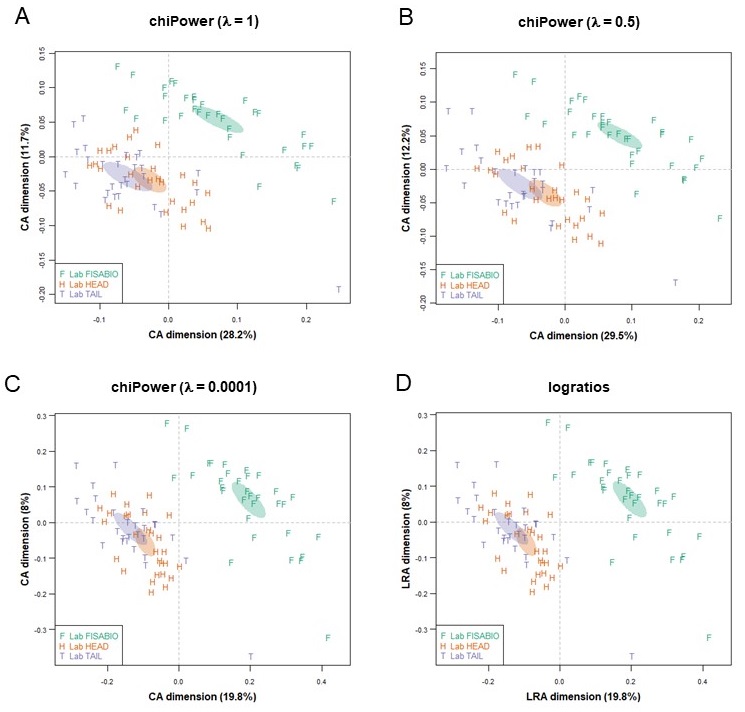}
\caption{Using the rabbits data set, three CAs, i.e. PCAs of chiPower-transformed compositions, with decreasing powers, and LRA analysis as the limit solution. A. The regular CA with power 1. B. CA with power 0.5 (square root). C. CA with power 0.0001. D. Logratio analysis (LRA). C and D are identical in their coordinate values to the fourth decimal. The ellipses are 95\% bootstrap confidence regions for the means of the three groups of points corresponding to three testing laboratories.}
\label{Rabbits_CA}
\end{center}
\end{figure}

To further illustrate the theoretical convergence of these geometries, Figure \ref{Rabbits_CA} shows the two-dimensional results of the CA for $\lambda = $1 (original CA), 0.5 (CA on square-root data), 0.0001 (CA on ten thousandth-root data), and finally LRA. 
As shown in Supplementary Material S3, these CAs are identical to PCAs on chiPowered data.
Figures \ref{Rabbits_CA}C and \ref{Rabbits_CA}D are identical in their coordinates up to four decimals --- the maximum absolute difference over all coordinate values is 0.00006.
The three groups of points correspond to three different laboratories which performed the testing, where it can be seen that one was quite different from the other two.

%-------------------------------------------------------------
\subsection{Unsupervised learning: compositions with zeros}
Here both the `Rabbits' and the `Crohn' data sets will be used to demonstrate how the chiPower transform can handle data zeros.
To simulate a situation where zeros are present in the `Rabbits' data, a count of 20 was temporarily regarded as the detection limit and all values less than 20 in the original matrix of microbial gene counts were set to 0.
This resulted in a data matrix with 25035 zeros, which is 7.1\% of the $89\times3937$ data matrix. 
This matrix was then closed to compositions, and analysed in a similar way as before.
In order to compare the results using the chiPower and logratio transformations, the zeros were imputed using the function \texttt{cmultRepl} in the \texttt{zCompositions} R package \citep{Palarea:15}, which is one of the popular ways of zero replacement.
The chiPower-transformed geometry of the data (with zeros) was then compared to the logratio geometry of the data matrix with zeros replaced.

\begin{figure*}[h]
\begin{center}
\includegraphics[width=12cm]{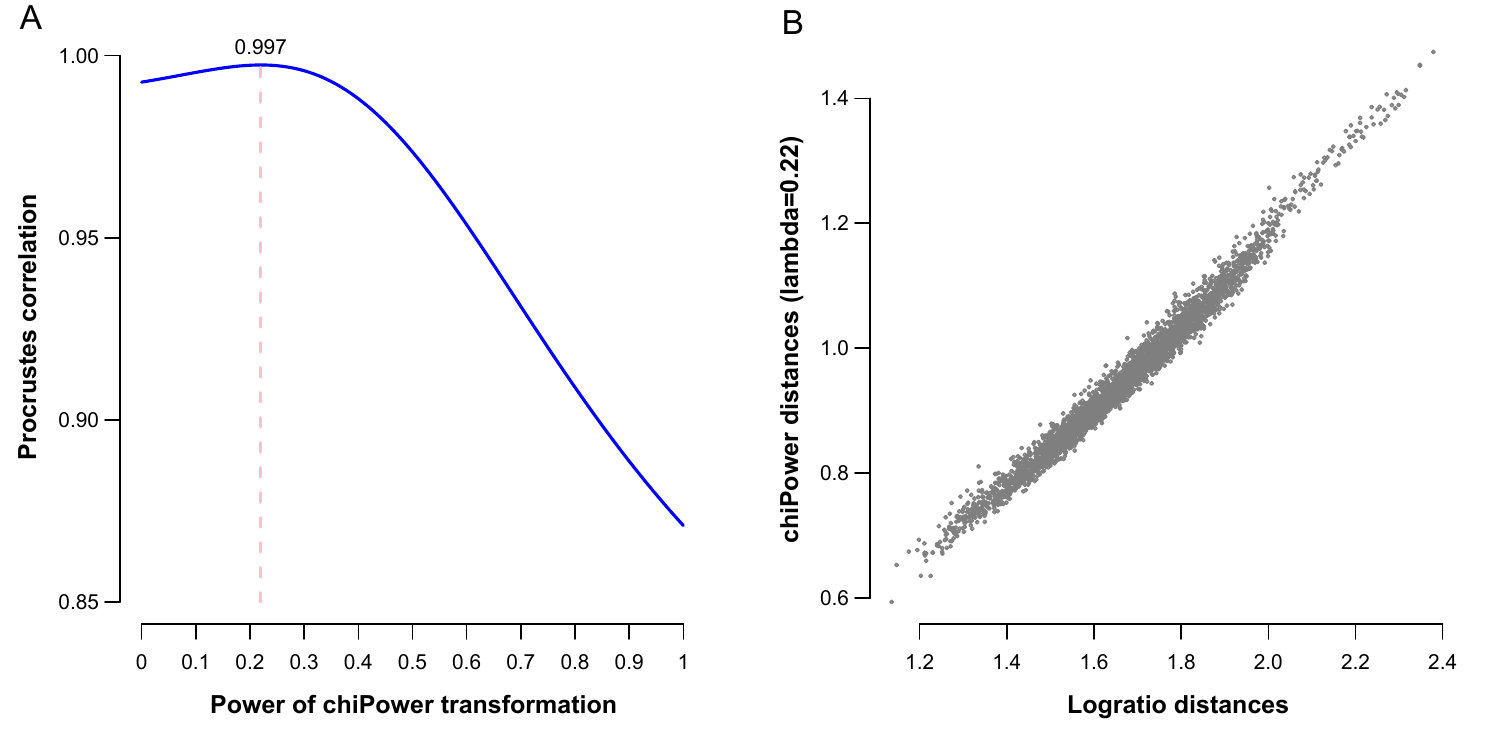}
\caption{A. The Procrustes correlations, measuring proximity to isometry between the exact logratio geometry and the geometry of correspondence analysis on compositional data subjected to Box-Cox transformation with different powers. The correlation is at an optimum value of 0.997 for a power of 0.22. B. In the respective 88-dimensional spaces, the chi-square distances are very similar to the logratio distances. }
\label{Rabbits_zeros}
\end{center}
\end{figure*}

As shown in Figure \ref{Rabbits_zeros}, the chiPower distances cannot reproduce exactly the logratio distances, because they are operating on slightly different data matrices, and thus convergence to logratio distances cannot be attained.
However, the geometries can come very close to each other depending on the power transformation selected.
Figure \ref{Rabbits_zeros}A shows that, as the power decreases, an optimal value of the Procrustes correlation is reached, equal to 0.997, at $\lambda = 0.22$, which is close to a fourth-root transformation.
The concordance of the chiPower and logratio distances can be seen in Figure \ref{Rabbits_zeros}B.
%Since we know the true values of the constructed zeros in the data, the comparison could also be made directly between the chiPower-transformed data and the original logratio transformed data. 
%The highest Procrustes correlation in this case is 0.990, only slightly less than 0.997 above, but for a power of $\lambda=0.59$, which is close to a square root transform.

\begin{figure}[h]
\begin{center}
\includegraphics[width=12cm]{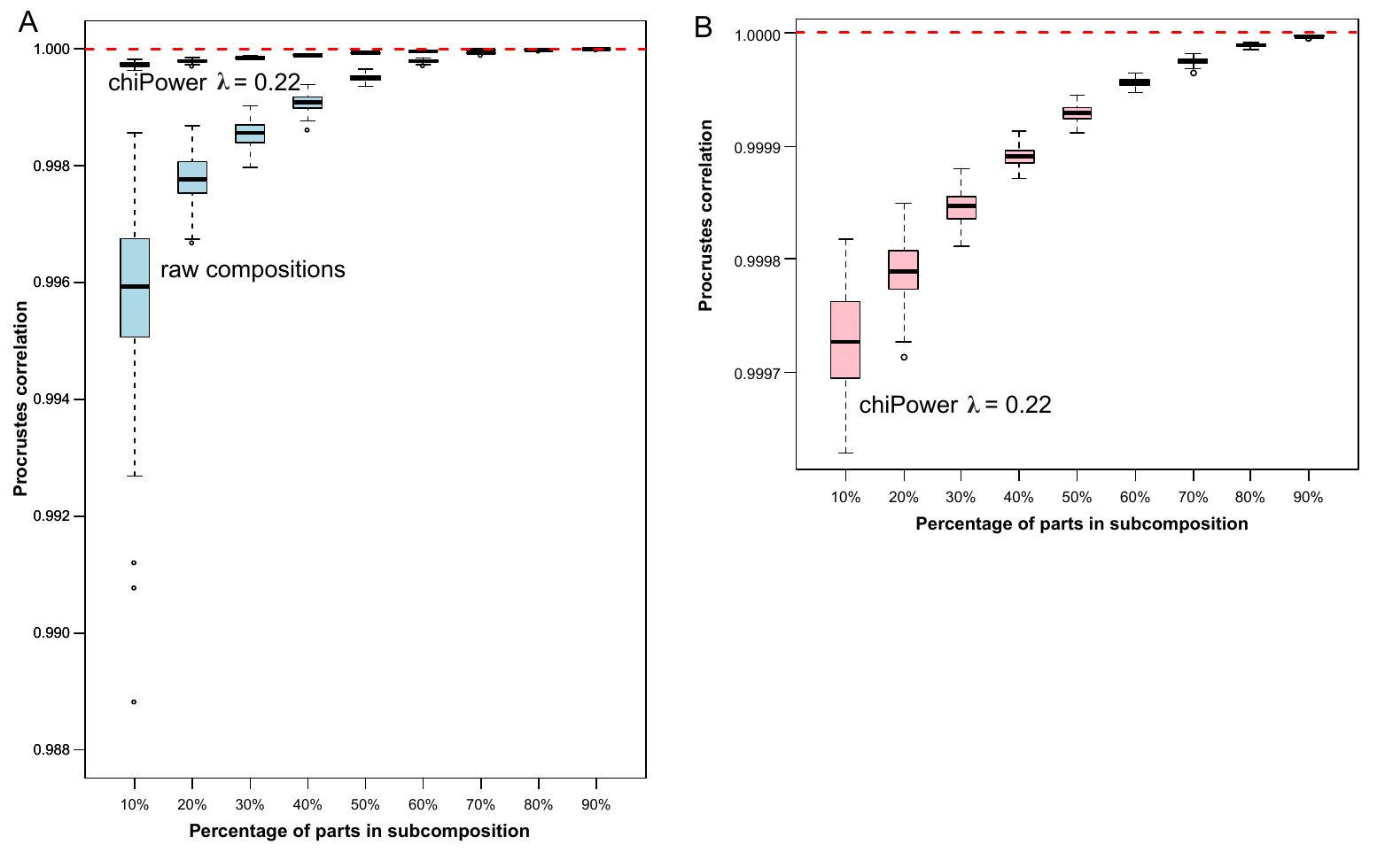}
\caption{Deviations from exact coherence (Procrustes correlation = 1) for 1000 random subcompositions of sizes 10\% up to 90\% of the 3937-part Rabbits data. For each size of subcomposition, a boxplot of the Procrustes correlations is shown, for the untransformed raw compositions and for chiPowered compositions. A. The lower sequence of boxplots is for the raw compositions, where no transformation is made at all, and deviations from coherence are much larger, especially for subcompositions with less parts. The upper sequence is for chiPower-transformed data, with power = 0.22 (the value obtained from the exercise on isometry), where deviations are very small, even in the smaller subcompositions. B. The same boxplots as the upper sequence in A., i.e. the chiPowered data, with expanded vertical scale.}
\label{Coherence_Rabbits}
\end{center}
\end{figure}

Since the chiPower-transformed data with $\lambda=0.22$ are close to isometry, it is expected that they will also be close to coherence. 
This is assessed by taking many random subcompositions, as described in Section 2.5, each of which is reclosed and its subcompositional part geometry compared with that of the corresponding subset of parts in the full composition. Once again, Procrustes correlation is used to measure the degree of coherence..
To contrast this with doing no change at all to the compositional data, the raw untransformed compositions were first assessed for isometry, which means that the regular Euclidean distance geometry on the raw compositions was correlated with the logratio geometry. 
The Procrustes correlation was computed as 0.891, and so it is expected that the coherence of the untransformed compositions will be worse than the quasi-isometric chiPower-transformed compositions.
This is indeed how it turns out in the subcompositional coherence exercise, shown in Figure \ref{Coherence_Rabbits}.

The same exercise was performed for the Crohn data set, and similarly successful results were obtained, given in Supplementary Material Section S4, where the optimal value of the power was $\lambda=0.25$.
%As explained in Section 2.1, this data set exists in two versions: the original data set, which contains zeros, and the modified data set, which is the original data set to which 1 has been added to every value. To validate the chiPower transformation, the original data set is chiPower-transformed, and its closeness to the logratio geometry of the modified data is measured.
%The result is an optimal power of $\lambda=0.25$, giving a Procrustes correlation of 0.902.
The result turns out to be dependent on the zero replacement.
%For example, if the zero counts are replaced using the function \texttt{cmultRepl} in package \texttt{zCompositions} \citep{Palarea:15}, which is a popular strategy in CoDA, the optimal power is $\lambda=0.18$, giving a Procrustes correlation of 0.948.
Supplementary Material Section S1 further investigates the effect of using different zero replacements, for example adding 0.5 to the original data, or simply substituting the zeros by 0.5.
%CA is useful once more, since the effect on the total variance of the data set can be judged using CA's total inertia measure, which is applicable to all the variations of the data set, including the original version with zeros.

%-------------------------------------------------------------
\subsection{Supervised learning: use of power transformations}
Compositions can serve as predictors of a response, or can form a multivariate response to other explanatory variables (e.g., \cite{Yoo:22}). In the latter case, isometry will still be relevant, since this affects the total compositional variance to be explained. 
Attention is restricted here to the former case,  where  the issue of isometry is no longer relevant but coherence certainly is, since the effect sizes and interpretation of the predictors should not depend on the particular (sub)composition they are part of -- see the next section.
Since there are many parts in a composition, the question of variable selection is first addressed in this section, comparing the predictors that are logratio transformed or chiPowered. 

The Crohn data set, with 975 samples and 48 bacteria, is used for this purpose since it has a dichotomous response $y = $ Crohn (patient with Crohn disease), or $y =$ no (no disease), to be predicted from the compositions.
Moreover, logistic regression models for predicting Crohn, using LRs, have already been fitted in two different ways, by \cite{Coenders:22} and \cite{Calle:23}.
\cite{Coenders:22} proposed three forward stepwise algorithms for choosing LRs, the first one being unrestricted choice from all possible LRs, of which there are $\frac{1}{2}\times 48\times 47 = 1128$. 
The available stopping criteria options were the Akaike information criterion (AIC), the stronger Bayesian information criterion (BIC) and the even stronger penalty on the number of variables in the model using the Bonferroni rule. 
This approach is implemented in the function \texttt {STEPR()} in the R package \texttt{easyCODA} \citep{Greenacre:18}. 
For the present application, the BIC stopping criterion will be used.

Using a different approach, \cite{Calle:23} includes all the LRs and imposes ElasticNet penalization on the predictors \citep{Hastie:09}, as implemented in the package \texttt{coda4microbiome}.

The above two approaches will be contrasted with simply using the power-transformed compositions, where the power is used as a tuning parameter to optimize the prediction.
This third option using chiPower is the only one of the three that uses the original version of the data with zeros.
Notice that the chi-square standardization and the multiplication by $\frac{1}{\lambda}$ in the chiPower transformation are not necessary here, as such scale changes do not affect the predictions, just the values of the regression coefficients.
Since \cite{Calle:23} uses the area under curve as a measure of prediction, and optimizes the variable selection using ten-fold cross-validation, the same approach is adopted here, to ensure comparability.
The results are summarized in Table \ref{Predictions}.

\smallskip
{\small
\setlength{\tabcolsep}{2pt}
\begin{center}
\begin{table*}[ht]
\begin{tabular}{lcccccccc}
\ & \multicolumn{4}{c}{\em Full data set} & \multicolumn{4}{c}{\em Crossvalidation}\\
 Method  &  LRs  & Parts & AUC  & Acc'y \qquad &  Sens'y & Spec'y & AUC & Acc'y\\[2pt]
\hline 
%\rule{0pt}{-0.5ex}\\
Forward stepwise &  11  & 19  &  0.859 & 82.2\% \qquad & 0.826 &  0.704 & 0.831 & 79.2\%\\
ElasticNet penalty &  27  & 24  &  0.848 & 80.2\% & NA &  NA & 0.824 & NA\\
Power ($\lambda=0.28$) &  --  & 14  &  0.859 & 81.6\% \qquad & 0.819  &  0.697 & 0.826 & 78.6\%\\
\hline\\
\end{tabular}
\caption{Results from three alternative ways of predicting Crohn's disease, based on different transformations of the compositional data: the first two using logratios on the modified version of the data set (with 1 added to all cells, from the \texttt{coda4microbiome} package), and the last using an optimized power transformation (power = 0.28) of the compositional parts from the original data set, with zeros. Different variable selection strategies are used: ElasticNet penalization, and the Bayesian information criterion (BIC) for the other two.
(LRs: the number of selected logratios. Parts: the number of parts. AUC: area under curve. Acc'y: accuracy (percent of correct predictions). Sens'y: sensitivity. Spec'y: specificity. NA: result not available in \cite{Calle:23})}
\label{Predictions}
\end{table*}
\end{center}}

The performance of all three is similar, but the simpler power transformation of the compositions needs only 14 parts. The ElasticNet approach \citep{Calle:23} chooses 27 logratios, involving 24 parts, while the forward stepwise approach \citep{Coenders:22} selects 11 logratios, involving 19 parts.
Ten-fold cross-validation, using the same folds, evaluates the performance of each approach. Since the cross-validation AUC of the ElasticNet approach is an average of the AUCs of the ten folds, the  mean AUC is also calculated for the other two methods.   
The power that is optimal in this supervised learning problem is $\lambda=0.28$, slightly higher than the power of 0.25 that was optimal in the unsupervised objective reported in Supplementary Material Section S5.
The question of coherence and interpretation of the results of this third approach is dealt with in Section 3.4.

\subsection{Coherence of the modelling with power-transformed compositions}

In the previous subsection, a small subset of 14 parts, power-transformed, was identified as good predictors of the Crohn disease response.
In this subsection the results and their interpretation are explained and it is investigated how the results would have changed if a subcomposition had been observed.
Such a subcomposition would include the selected 14 parts, but would have different compositional values due to the closing of the subcomposition.
For predictors in the form of LRs or ALRs, their exact coherence ensures that the results remain the same -- that is, a result for any subcomposition would be identical if any number of compositional parts were eliminated (or added) to the data set and the data reclosed to sum to 1.
But for other transformations such as the present power-transformed one, a check is necessary on the extent of the lack of coherence in the results.

\begin{figure}[b]
\begin{center}
\includegraphics[width=10cm]{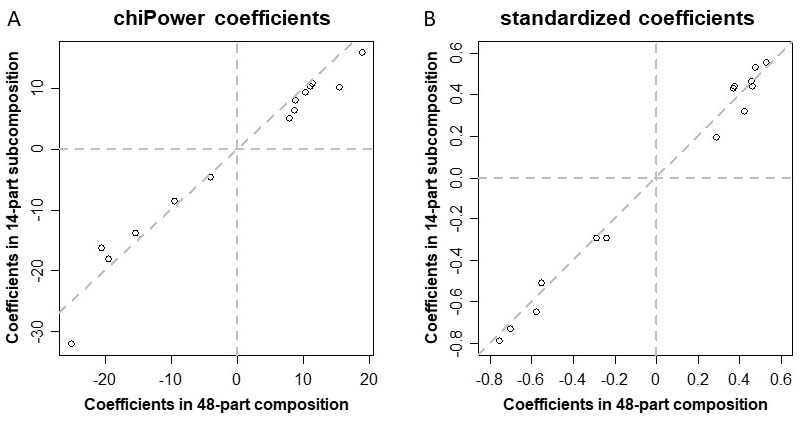}
\caption{Scatterplots of regression coefficients from models of same 14 predictors, fitted to data from the ``full" 48-part composition, and to data from the 14-part re-closed subcomposition. A. Coefficients of chiPower transformed variables ($\lambda=0.28$) in each case. B. Standardized regression coefficients from each model. In both scatterplots the aspect ratio is 1 and the diagonal line at 45 degrees represents exact proportionality. }
\label{Coherence}
\end{center}
\end{figure}

The first check is to isolate the 14 parts in a subcomposition, reclose, and then repeat the model.
Because of the change of scale of the compositional values, the coefficients change (Figure \ref{Coherence}A) but are still almost proportional to one another. 
Notice that the variables in this case are chiPower transformed, so the chi-square standardization is incorporated.
For predictors that are standardized in each case (i.e., mean 0, variance 1) in order to obtain standardized regression coefficients, which gives identical results for chiPower or simply power transformed variables, the standardized coefficients are compared in Figure \ref{Coherence}B and are seen to be in the same order and practically the same.
The concordance between the two sets of coefficients is clear, and in this sense the modelling is very close to coherence.
Compared to the optimized results of Table \ref{Predictions}, the AUC and accuracy, when the model is fitted to the closed 14-part subcomposition, both drop slightly from 0.859 to 0.847 and from 81.6\% to 79.7\%, respectively.
This loss of predictivity might well be improved if the power was tuned specifically to optimizing coherence in the modelling, as opposed to the unsupervised objective of optimizing the isometry with respect to the sample logratio geometry.

\begin{figure}[ht]
\begin{center}
\includegraphics[width=12.4cm]{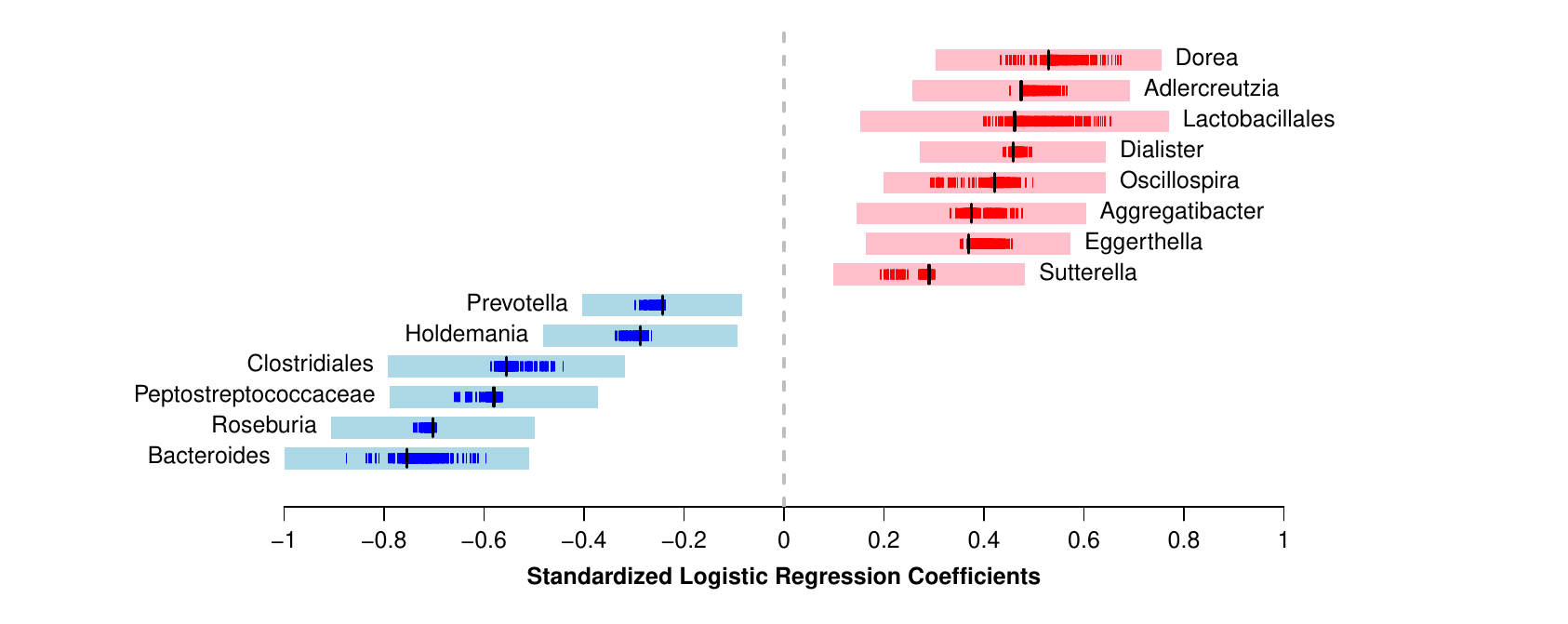}
\caption{Standardized logistic regression coefficients from 1000 analyses using different subcompositions of the Crohn data set. Each subcomposition contains the 14 parts used in the original model reported in the third row of Table \ref{Predictions} with random extents (1 to 33 additional parts), then each data set is closed and the parts are power-transformed using power = 0.28.  The pale red and blue bars show 95\% confidence intervals for the estimated coefficients in the original model and the vertical black lines are the point estimates, at the midpoints of the confidence intervals. For each variable there are 1000 vertical red or blue lines showing the estimates from the models using subcompositions. Positive coefficients increase the log-odds to Crohn's disease, negative coefficients decrease the log-odds.}
\label{Subcompositions}
\end{center}
\end{figure}

To further investigate the coherence issue in the modelling, random subcompositions involving the same 14 parts but also additional parts, randomly selected and of random extents (from 1 to 33 additional parts), are added to the data set. 
For each of these, the subcomposition is closed and then power-transformed using $\lambda=0.28$ (Table \ref{Predictions}), and the logistic regression repeated using the 14 parts as predictors.
Figure \ref{Subcompositions} shows the results for 1000 random subcompositions.
The original coefficients are shown as vertical black lines, in the centre of a 95\% confidence interval in pink, according to the margin of error $\pm 1.96\,\textrm{SE}$ for each coefficient.
The estimated coefficients in the subcompositions are shown as vertical red or blue lines (for positive and negative coefficients, respectively), where it can be seen that they all span the original estimates, and are well within the confidence intervals.
For each of the 1000 subcompositions the accuracies and AUCs are also computed, and 95\% of the accuracies are between 80.0\% and 81.4\%, while 95\% of the AUCs are between 0.848 and 0.859.
This further demonstrates that the logistic regression results would be substantively the same for any subcomposition, so that the modelling using power-transformed compositions is coherent for all practical purposes, further supporting the good performance of these power-transformed predictors in Table \ref{Predictions}.  

\begin{figure}[ht]
\begin{center}
\includegraphics[width=8.5cm]{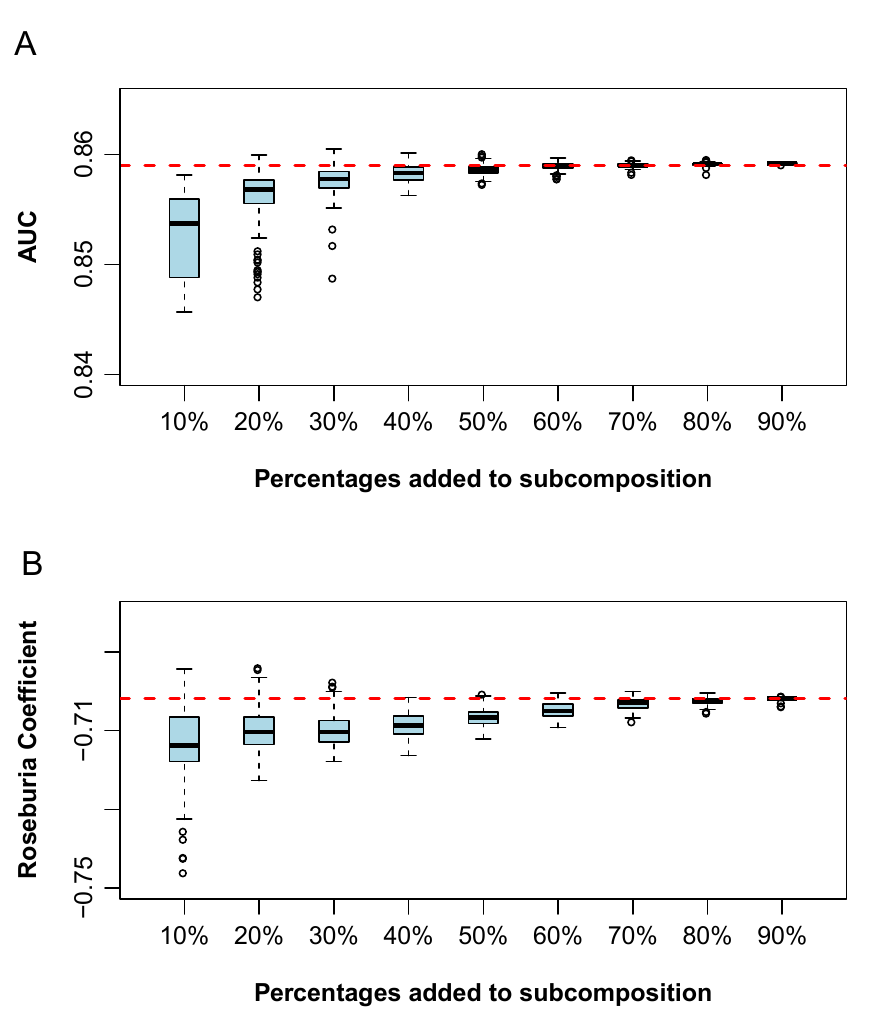}
\caption{Results from logistic regressions performed on 100 subcompositions for each of nine different percentages of the remaining parts added to the basic set of 14 power-transformed parts that were used in the original model predicting Crohn's disease. Boxplots show the dispersions for each subcomposition of increasing sizes. A. AUC of prediction. B. Standardized regression coefficients for Roseburia. The red dashed lines show the original values in the full composition.}
\label{SubcompositionSizes}
\end{center}
\end{figure}

Another diagnostic of coherence is to see how much the regression coefficients change as a function of the sizes of the chosen subcompositions.
Random subcompositions of 10\%, 20\%, etc., up to 90\% of the 33 remaining microbial taxa were taken, where the 14 parts in the original model are always included, 100 subcompositions in each case. 
The dispersions of the AUC values (original value of $0.859$ in the model -- see Table \ref{Predictions}) and for the regression coefficients of Roseburia (original standardized coefficient in the model equal to $-0.702$) are shown in Figure \ref{SubcompositionSizes}, in the form of boxplots.
For small subcompositions the AUCs are under-estimating, as already seen when just the 14-part subcomposition (with no others added) was analyzed.  The standardized coefficients of Roseburia are more negative but both the model AUCs and these coefficients converge to the values in the original model as the subcompositional size increases. 
The dispersion of these coefficients should be judged against the margins of error of the estimates in the full composition. 
For example, Roseburia's coefficient estimate is $-0.702$, with a SE of $0.104$, giving a 95\% confidence interval of [ $-0.906$, $-0.498$ ], much wider than the dispersions shown in Figure \ref{SubcompositionSizes}B.

As for the interpretation, this is made directly on the part values (power-transformed), not on logratios, which is a considerable simplification. 
The standardized regression coefficients, shown in Figure \ref{Subcompositions}, give a model for log-odds of Crohn's disease in terms of the 14 standardized predictors of the following form, showing just the extreme negative and positive terms:
\begin{equation}
  \begin{split}
   \log(\frac{p}{1-p}) = 1.238 &- 0.754\,{\small\rm Bacteriode}^* 
                               - 0.702\,{\small\rm Roseburia}^*
                               \, \cdots \, \\
                               &+ 0.474\,{\small\rm Adlercreutzia}^*
                               + 0.529\,{\small\rm Dorea}^*
  \end{split}
\end{equation}
where $^*$ indicates the standardized power-transformed variables.
Alternatively, the equivalent model can be expressed in terms of the values in the original composition using power-transformed variables and no standardization, where the magnitude and order of the coefficients changes according to the ranges of the different predictors:
\begin{equation}
  \begin{split}
   \log(\frac{p}{1-p}) = 3.197 &- 5.027\,{\small\rm Roseburia}^{0.28} 
                               - 4.866\,{\small\rm Peptostreptococcaceae}^{0.28}
                               \, \cdots \, \\
                               &+ 5.208\,{\small\rm Eggerthella}^{0.28}
                               + 7.948\,{\small\rm Adlercreutzia}^{0.28}
  \end{split}
  \label{RegressionModel}
\end{equation}

Whichever form is reported, it should be remembered that the effect sizes are applicable to infinitesimal (i.e., very small) changes in the predictors, and should not be taken as linear effects as in regular regression.
Like partial derivatives, these are measures of local changes.
This is due to a change in one compositional value affecting all the others.
For example, suppose all the predictors are at their mean values. The value of the regression equation in (\ref{RegressionModel}), including the constant is computed to be 2.583, back-transformed to a probability  of Crohn's disease equal to 0.930 ($p = e^{2.583}/(1+e^{2.583}) = 0.930$).
Suppose the compositional mean value of Roseburia is multiplied by 20, which is still within the range of this bacteria's observed values. 
Simply making this increment and applying the model to the new set of power-transformed values results in the value 0.667. This back-transforms to a probability of 0.661, less than 0.930, as expected since Roseburia's coefficient in the regression is negative.

But one cannot simply change a compositional value as one would do with regular statistical variables, since the other compositional values are affected by the change.
Hence, the increased value of Roseburia has to be compensated by a decrease in the compositional values of the other bacteria. 
Applying a proportional decrease to the other bacteria to obtain a composition that sums to 1 and applying the model formula leads now to a value of 0.607 and a back-transformed probability of 0.647, which would be a more accurate estimate of the effect of the Roseburia increase.

This issue of quantifying the correct effect sizes due to the nature of the compositional data, taking into account that a change in one part affects the others, is similarly present when logratios are used as predictors and the model is expressed as a log-contrast \citep{Coenders:22}.

%-------------------

\section{Conclusion}
This paper demonstrates that an alternative pipeline is possible for analysing compositional data, using the chiPower transformation. This transformation combines a Box-Cox style of power transformation with the chi-square standardization that is inherent in correspondence analysis.
The choice of the power gives the approach its flexibility.
Unlike logratio transformations, this transform allows data zeros -- notice that in the Crohn application, 28.8\% of the original data matrix are zeros and need replacement in order to compute logratios.
In an unsupervised learning context, where understanding of the data structure is sought, the power can be identified to maximize the proximity of the sample geometry of chiPower transformed compositions to the sample logratio geometry, using the Procrustes correlation as a measure of the closeness to isometry.
In a similar way, the Procrustes correlation between the geometries of subsets of parts in a composition and the same parts in subcompositions gives a quantitative assessement of (subcompositional) coherence. 
For supervised learning where the compositions are predictors of a response, the power serves as a tuning parameter to optimize prediction of the response, preferably using cross-validation.  
In this case, where a subset of power-transformed predictors is selected and a model fitted, the coherence can be assessed by repeating the model fitting on subcompositions of different sizes and observing how the model estimates are affected.

Overall, in summary, the chiPower transformation, supported by diagnostics to assess the properties of isometry and coherence, can present a simpler and more easily interpretable alternative to the logratio transformation, with the great advantage that no zeros need replacing.

These results give food for thought about the role that logratios play in compositional data analysis.
When data are all positive, the logratio approach can be adopted, with its favourable property of exact coherence.
But when there are data zeros, the user has a choice, either to use an algorithm to literally create data for the sake of using logratios, or use an alternative approach that needs no change to the data and that can be shown to be almost isometric and coherent in terms of the research objective, unsupervised or supervised. 
The different zero replacement methods can lead to different results (see Supplementary Material Section S1) and there is apparently no clear consensus about which is preferred in a specific context, and the choice of method may partially depend on the nature of the zeros.
Hence, it may be that an alternative approach, such as the chiPower transformation presented here, is preferable in the case of data zeros, especially many data zeros. 
Note that the measure of coherence can be determined without zero replacement.

In summary, transformations such as chiPower, which are highly coherent and needing no zero replacement, are proposed as a preferred first choice for analysing compositional data that have zeros.
Then, if logratios are of interest, for whatever reason, the zero replacement method can be chosen that leads to logratio-transformed data that come closest (for example, in terms of isometry or model accuracy) to the data transformed by the preferred method. 
For strictly positive data, both approaches are possible, the purely logratio approach, where the final interpretation is in terms of logratios, or the chiPower approach where the interpretation is in terms of the original compositional parts, which may be easier for the practitioner.

%%%%%%%%%%%%%%%%%%%%%%%%%%%%%%%%%%%%%%%%%%%%%%%%%%%%%%%%%%%%%%%%%%
%%%%%%%%%%%%%%%%%%%%%%%%%%%%%%%%%%%%%%%%%%%%%%%%%%%%%%%%%%%%%%%%%%

\section*{Appendix 1}
\leftline{\large \bf Relationship between chiPower and CLR transformations}

\medskip
\noindent
The fact that the chiPower transformation links directly to LRA, which is the PCA of the CLR-transformed positive compositional data, implies a direct link from the chiPower transform and the CLR transform.
To show this, first consider this result, for the positive composition ${\bf x}_i = [\  x_{i1}\ x_{i2}\ \cdots \ x_{iJ} \ ]$ in the $i$-th row of the compositional data matrix $\bf X$.
Let $y_{ij}{\scriptstyle [\lambda]} = x_{ij}^\lambda / \sum_k x_{ik}^\lambda$, i.e. the closed powered compositions.
The convergence of $J y_{ij}{\scriptstyle [\lambda]}$ to the CLR transformation, in Box-Cox formulation, is as follows
\begin{equation}
    \lim_{\lambda\to 0} \frac{1}{\lambda} \left( J \frac{x_{ij}^\lambda}{\sum_k x_{ik}^\lambda} - 1 \right) = {\rm CLR}({\bf X})_{ij} = \log\left(\frac{x_{ij}}{g({\bf x}_i)}\right)
    \label{lim_to_CLR}
\end{equation}
where $g({\bf x}_i)$ is the geometric mean of the $J$ elements of ${\bf x}_i$.
To show this, divide the numerator $x_{ij}^\lambda$ and denominator $\sum_j x_{ij}^\lambda$ both by $g({\bf x}_i)^\lambda$.
\begin{equation}
    \lim_{\lambda\to 0} \frac{1}{\lambda} \left( J \frac{x_{ij}^\lambda / g({\bf x}_i)^\lambda}{\sum_k x_{ik}^\lambda / g({\bf x}_i)^\lambda} - 1 \right)
    \label{lim_to_CLR_with_g}
\end{equation}
In the denominator $\lim_{\lambda\to 0} \sum_k x_{ik}^\lambda = J$ and $\lim_{\lambda\to 0} g({\bf x}_i)^\lambda = 1$, hence the limit reduces to
\begin{equation}
    \lim_{\lambda\to 0} \frac{1}{\lambda} \left( \left[{\frac{x_{ij}}{g({\bf x}_i)}}\right]^\lambda - 1\right) = \log\left(\frac{x_{ij}}{g({\bf x}_i)}\right)
    \label{lim_to_CLR2}
\end{equation}
using the Box-Cox theorem.
A different proof of this result is given in the Appendix of \citep{Tsagris:16} using series expansions.

To prove the convergence properties of the chiPower transform, Tsagris el al's style of proof will be used here. 
The following results are used in the proof, both based on Taylor series expansions of these functions of $\lambda$, around the value $\lambda=0$:
\newenvironment{itemize*}%
  {\begin{itemize}%
    \setlength{\itemsep}{5pt}%
    \setlength{\parskip}{2pt}}%
  {\end{itemize}}
\hspace{1cm}  
\begin{itemize*}
    \itemindent=30pt
    \item  $x^\lambda = 1 + \lambda \log(x)  + O(\lambda^2)$
    \item  $(1+\lambda x)^{a} = 1+a\lambda x +  O(\lambda^2)$
\end{itemize*}
In the proof, the terms in $O(\lambda^2)$ (including higher power) are written just the first time they occur in an expansion and then omitted since they will eventually disappear in the limit.

The basic chiPower transformation is $y_{ij}{\scriptstyle [\lambda]} / \sqrt{\bar{y}_j{\scriptstyle [\lambda]}}$ where $\bar{y}_j{\scriptstyle [\lambda]} = (1/I) \sum_i y_{ij}{\scriptstyle [\lambda]}$, the column means of the $y_{ij}{\scriptstyle [\lambda]}$. The numerator and denominator are first handled separately.

The numerator is expanded as follows:
\begin{align}
  y_{ij}{\scriptstyle [\lambda]} &= \frac{x_{ij}^\lambda}{\sum_k x_{ik}^\lambda} \nonumber \\
         &= \frac{1 + \lambda \log(x_{ij}) + O(\lambda^2)}
                 {\sum_k (1 + \lambda \log(x_{ik}) + O(\lambda^2)} \nonumber\\
         &= \Big(1 + \lambda \log(x_{ij})\Big) \frac{1}{J} \Big(1 + \frac{\lambda}{J} \sum_k \log(x_{ik}) \Big)^{-1} \nonumber \\
         &= \frac{1}{J} \Big(1 + \lambda \log(x_{ij})\Big) \Big(1 - \frac{\lambda}{J} \sum_k \log(x_{ik}) +O(\lambda^2)\Big) \nonumber\\
         &= \frac{1}{J} \Big(1 + \lambda \log(x_{ij}) - \frac{\lambda}{J} \sum_k \log(x_{ik}) + O(\lambda^2)\Big) \label{numerator}
\end{align}

\noindent
From this result the inverse of the denominator is expanded as
\begin{align}
  1/\sqrt{\bar{y}_j{\scriptstyle [\lambda]}} &= \Big(\frac{1}{I} \sum_i y_{ik}{\scriptstyle [\lambda]}^\lambda\Big)^{-0.5} \nonumber\\
         &= I^\frac{1}{2}J^\frac{1}{2} \Bigg(\sum_i  \Big(1 + \lambda \log(x_{ij}) - \frac{\lambda}{J} \sum_k \log(x_{ik}) + O(\lambda^2)\Big)\Bigg)^{-0.5} \nonumber\\
         &= J^\frac{1}{2} \Bigg(1 + \frac{1}{I}\sum_i \Big(\lambda \log(x_{ij}) - \frac{\lambda}{J} \sum_k \log(x_{ik}) + O(\lambda^2)\Big)\Bigg)^{-0.5} \nonumber \\        
         &= J^\frac{1}{2} \Bigg(1 + (-0.5)\frac{1}{I}\sum_i \Big(\lambda \log(x_{ij}) - \frac{\lambda}{J} \sum_k \log(x_{ik}) + O(\lambda^2)\Big)\Bigg) \label{denominator}
\end{align}

\noindent
The division of numerator by denominator is thus the product of (\ref{numerator}) and (\ref{denominator}). Many products of terms are $O(\lambda^2)$ and the only ones remaining are those that are multiplied by the 1's in each bracket, reducing to
\begin{multline}
    \frac{1}{J^\frac{1}{2}} \Big(1 + \lambda \log(x_{ij}) - \frac{\lambda}{J} \sum_k \log(x_{ik})  \\
    - 0.5\frac{1}{I}\sum_i \Big(\lambda \log(x_{ij}) - \frac{\lambda}{J} \sum_k \log(x_{ik}) + O(\lambda^2) \Big)    
\end{multline}
%\vspace{-0.3cm}
so that 
\begin{equation}
    \frac{1}{\lambda} \left( J^\frac{1}{2} \frac{y_{ij}{\scriptstyle [\lambda]}}{\sqrt{\bar{y}_j{\scriptstyle [\lambda]}}} - 1\right)  \rightarrow {\rm CLR}({\bf X})_{ij} - 0.5 \ \overline{{\rm{CLR}({\bf X})}}_j \ \ {\rm as} \ \lambda \rightarrow 0
    \label{limit}
\end{equation}
where ${\rm CLR}({\bf X})_{ij} = \log(x_{ij}) - \frac{1}{J} \sum_k \log(x_{ik})$ is the centered logratio defined in (\ref{lim_to_CLR}) and $\overline{{\rm{CLR}({\bf X})}}_j$ is the $j$-th column mean of ${\rm CLR}({\bf X})$.

Thus, the limit is the CLRs shifted negatively by half their column means. 
Since the column means of this limit in (\ref{limit}) are equal to (plus) half the column means, the negative shift can be cancelled by a translation that adds half the column means of the transformation $(1/\lambda)\big(J^\frac{1}{2} y_{ij}{\scriptstyle [\lambda]} / \sqrt{\bar{y}_j{\scriptstyle [\lambda]}} - 1 \big)$.
This ``translated" version that converges to the CLR will be the default in the \textsf{R} function \texttt{chiPower}, but the ``unadjusted" version can also be obtained as an option. 
Of course, these options that shift each part by a constant amount make no difference to computing distances, covariances, or models, since the column means are eliminated or just affect the constant terms in models.
 Notice that the similar proof by \citep{Choulakian:23} is not for the chiPower transformation but for $y_{ij}{\scriptstyle [\lambda]}/\bar{y}_j{\scriptstyle [\lambda]}$, that is, dividing by $\bar{y}_j{\scriptstyle [\lambda]}$ rather than by $\sqrt{\bar{y}_j{\scriptstyle [\lambda]}}$. This ratio is a scalar multiple of the Pearson contingency ratio \citep{Greenacre:10a} and converges to centered CLRs since the $-0.5$ in (\ref{limit}) above becomes $-1$ and the $J^{1/2}$ is eliminated, giving the result:
 \begin{equation}
    \frac{1}{\lambda} \left( \frac{y_{ij}{\scriptstyle [\lambda]}}{\bar{y}_j{\scriptstyle [\lambda]}} - 1\right)  \rightarrow {\rm CLR}({\bf X})_{ij} -  \ \overline{{\rm{CLR}({\bf X})}}_j \ \ {\rm as} \ \lambda \rightarrow 0
    \label{limit2}
\end{equation}
In fact, it is clear from the proof in equations (\ref{numerator})--(\ref{limit}) that the general result for any power $\phi$ in the division $y_{ij}{\scriptstyle [\lambda]} / \bar{y}_j{\scriptstyle [\lambda]}^\phi$ can be obtained as follows:
 \begin{equation}
    \frac{1}{\lambda} \big( J^{1-\phi}\frac{y_{ij}{\scriptstyle [\lambda]}}{\left(\bar{y}_j{\scriptstyle [\lambda]}\right)^\phi} - 1\big)  \rightarrow {\rm CLR}({\bf X})_{ij} -  \phi \ \overline{{\rm{CLR}({\bf X})}}_j \ \ {\rm as} \ \lambda \rightarrow 0
    \label{limit3}
\end{equation}
of which (\ref{limit}) and (\ref{limit2}) are special cases for $\phi=$ 0.5 and 1 respectively.

These results are illustrated in the following R code, applied to the modified Crohn data without zeros. 

\smallskip

{\small
\begin{verbatim}
### The chiPower transformation
chiPower <- function(X, close=TRUE, power=1, chi=TRUE, 
                     BoxCox=TRUE, CLR=TRUE) 
{
#  X: the compositional data matrix (it is closed in case)
#  close: close the data after powering
#  power: power of the transformation
#  chi: apply chi-square standardization
#  BoxCox: apply Box-Cox style of transformation
#  CLR: translate columns so that convergence is to CLR transformation 
   foo <- as.matrix(X)
   foo <- foo / rowSums(foo)
   foo <- foo^power
   if(close)   foo <- foo / rowSums(foo)
   if(chi)     foo <- sweep(foo, 2, sqrt(colMeans(foo)), FUN="/")
   if(BoxCox & !chi)  foo <- (1/power)*(ncol(X)*foo - 1)
   if(BoxCox & chi)   foo <- (1/power)*(sqrt(ncol(X))*foo - 1)
   if (BoxCox & chi & CLR) foo <- foo + rep(colMeans(foo), each=nrow(X))
   X.chiPower <- foo
   X.chiPower
}

### Use modified Crohn data without zeros, Crohn1
library(easyCODA)
X <- CLOSE(Crohn1)

### CLRs
X.CLR <- CLR(X, weight=FALSE)$LR
### First 6 rows and 3 columns of CLR matrix
head(X.CLR,1:3])
#                 g__Turicibacter g__Parabacteroides g___Ruminococcus_
# 1939.SKBTI.0175       -4.920831          2.1617171          2.801403
# 1939.SKBTI.1068       -3.176255         -0.3430419          4.380173
# 1939.SKBTI047         -2.686822          1.2384461          2.980757
# 1939.SKBTI051         -1.956917          1.9142845          3.956586
# 1939.SKBTI063         -3.719167         -0.8287950          4.228865
# 1939.SKBTI072         -2.683135          6.1129017          2.948077

### chiPower without chi-square standardization and confirming (7)
### (compare these values with the CLR values given above)
lambda <- 0.001
head(chiPower(X, power=lambda, chi=FALSE))[,1:3]
#                 g__Turicibacter g__Parabacteroides g___Ruminococcus_
# 1939.SKBTI.0175       -4.914989          2.1577662          2.799038
# 1939.SKBTI.1068       -3.175177         -0.3469550          4.385789
# 1939.SKBTI047         -2.686208          1.2362093          2.982195
# 1939.SKBTI051         -1.957958          1.9131512          3.961451
# 1939.SKBTI063         -3.716163         -0.8323668          4.233885
# 1939.SKBTI072         -2.683284          6.1278452          2.948661

### chiPower with chi-square standardization but no CLR shift adjustment
head(chiPower(X, power=lambda, CLR=FALSE))[,1:3]
#                 g__Turicibacter g__Parabacteroides g___Ruminococcus_
# 1939.SKBTI.0175      -3.6753284          1.2511554          1.336367
# 1939.SKBTI.1068      -1.9333492         -1.2512999          2.920804
# 1939.SKBTI047        -1.4437711          0.3304323          1.519257
# 1939.SKBTI051        -0.7146141          1.0067617          2.497085
# 1939.SKBTI063        -2.4750092         -1.7362725          2.769121
# 1939.SKBTI072        -1.4408430          5.2176429          1.485772

### plot showing that it's just a shift away from CLRs (Figure 8)
plot(X.CLR$LR, chiPower.new(X, power=lambda, CLR=FALSE), 
     col=rep(rainbow(ncol(X)), each=nrow(X)))

### chiPower with chi-square standardization and CLR shift adjustment
### (this is the default of the chiPower funcion)
head(chiPower(X, power=lambda))[,1:3]
#                 g__Turicibacter g__Parabacteroides g___Ruminococcus_
# 1939.SKBTI.0175       -4.919562          2.1566333          2.797086
# 1939.SKBTI.1068       -3.177583         -0.3458220          4.381522
# 1939.SKBTI047         -2.688004          1.2359101          2.979976
# 1939.SKBTI051         -1.958847          1.9122396          3.957804
# 1939.SKBTI063         -3.719243         -0.8307946          4.229840
# 1939.SKBTI072         -2.685076          6.1231208          2.946490
\end{verbatim}
}

\medskip

\begin{figure}[h!]
\begin{center}
\includegraphics[width=6.5cm]{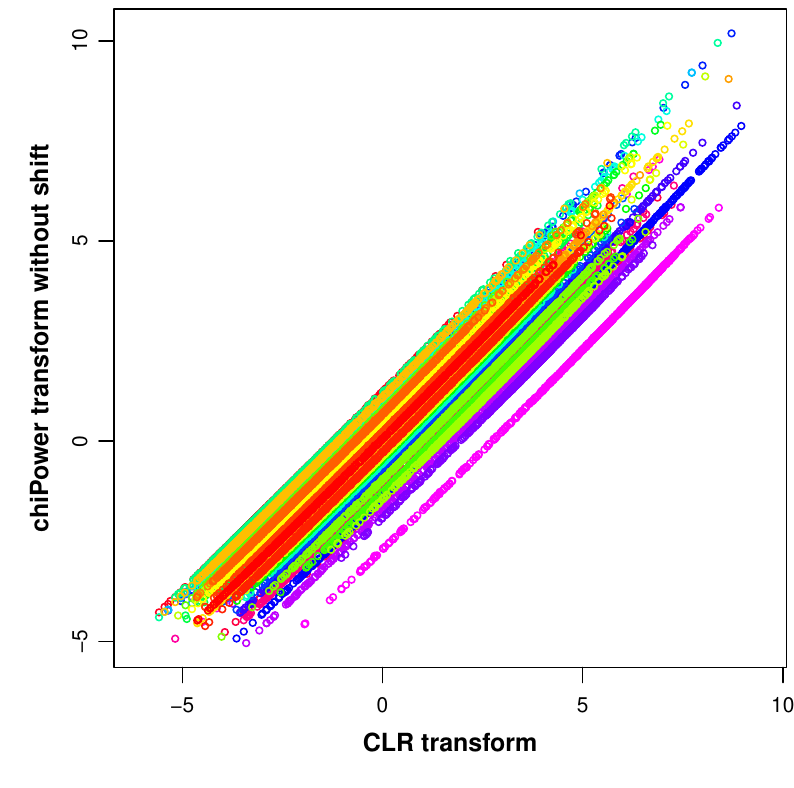}
\caption{Scatterplot of unadjusted chiPower transformation and the CLR transformation, showing that each part is a shift away from the CLR. This can easily be corrected if required, as in the next Figure \ref{chiPower_vs_CLR}. Each colour corresponds to one of the 48 compositional parts.} 
\label{CLR_without_shift}
\end{center}
\end{figure}

\begin{figure}[h!]
\begin{center}
\includegraphics[width=9cm]{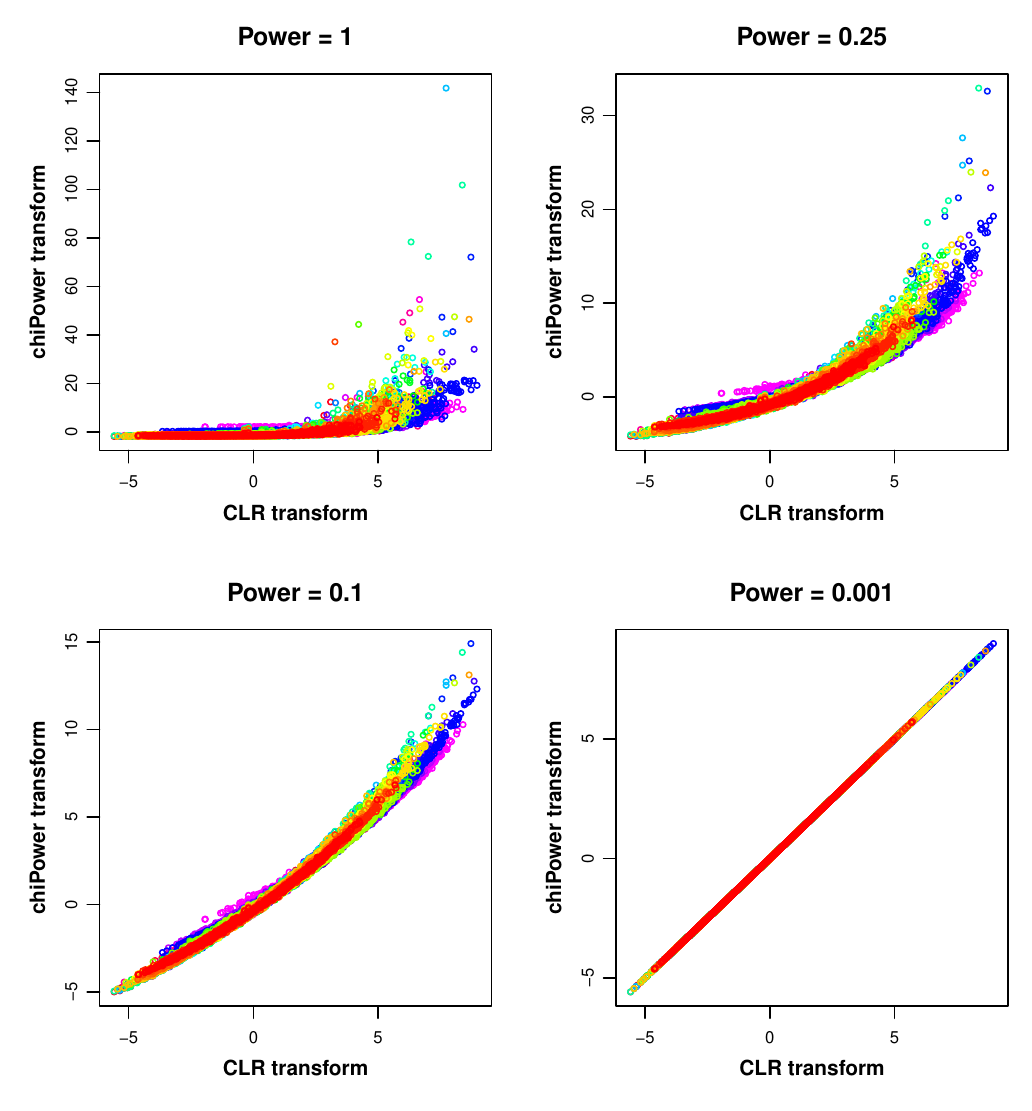}
\caption{Scatterplots of chiPower transformations for decreasing powers and the CLR transformation, with shift adjustment. Each colour corresponds to one of the 48 compositional parts} 
\label{chiPower_vs_CLR}
\end{center}
\end{figure}

\noindent
Figure \ref{CLR_without_shift} shows the scatterplot in the above code, for power $\lambda = 0.001$.  
Figure \ref{chiPower_vs_CLR} shows the comparisons of the CLR transformation (horizontal axis on each plot), and the chiPower transformation, with the shift adjustment, for decreasing powers 1, 0.25, 0.1 and 0.001.

\bigskip

\bigskip

\newpage

%end red
  
%%%%%%%%%%%%%%%%%%%%%%%%%%%%%%%%%%%%%%%%%%%%%%%%%%%%%%%%%%%%%%%%%%%%%%%%%%%%%%%%%%
\section*{Appendix 2: the Procrustes correlation}
\noindent
Procrustes analysis \citep{Gower:04} is a method for matching two multidimensional configurations by introducing translation, rotation and scaling operations to make them as similar as possible to each other.
It is used here to measure how similar two data structures are \citep{Krzanowski:87}, for example between the matrix ${\bf F}_1$ of principal coordinates from an LRA and the matrix ${\bf F}_2$ of principal coordinates from a PCA of a chiPower-transformed compositional data matrix.
Both ${\bf F}_1$ and ${\bf F}_2$ are assumed to have already been column-centred, which takes care of the translation operation, since this makes their sample means identically equal to the zero vector.
The first step is then to normalize the two configurations so they both have sums of squares equal to 1, which takes care of the scaling.
It just remains to find the rotation of one configuration to agree as closely as possible with the other, in the sense of least-squared differences between them, which is where the SVD is used.
Procrustes analysis and the computation of the Procrustes correlation proceed as follows. 
\vspace{-0.1cm}
\begin{flalign}
\nonumber & \textrm{1.\ \ Normalize both matrices:%\footnote{The notation ``trace" denotes the trace of a matrix (the sum of its diagonal values), and tr(${\bf A}\tr{\bf A}$) is the sum of squares of the elements of {\bf A}.}
} 
& \ & \hspace{-0.1cm}
\mkern-9mu {\bf F}_1^\ast = {\bf F}_1 / \sqrt{{\rm trace}({\bf F}_1\mkern-3mu\tr {\bf F}_1)} & \\
 & \ & \ & \hspace{-0.1cm} 
\mkern-9mu {\bf F}_2^\ast = {\bf F}_2 / \sqrt{{\rm trace}({\bf F}_2\mkern-3mu\tr {\bf F}_2)}& \\
& \textrm{2.\ \  Compute cross-products:}  & \ & \hspace{-0.1cm}
{\bf S}  = {\bf F}_1^{\ast{\sf T}} {\bf F}_2^\ast& \\[3pt]
& \textrm{3.\ \  Perform SVD:} & \ & \hspace{-0.1cm}
{\bf S}  = {\bf UD}_\alpha{\bf V}\tr & \\[3pt]
& \textrm{4.\ \  Rotation matrix:} & \ & \hspace{-0.1cm}
\mkern-2mu {\bf Q}  = {\bf V}{\bf U}\tr &  %\\
\end{flalign}
%\footnotetext {The notation ``trace" denotes the trace of a matrix (the sum of its diagonal values). The diagonal elements of ${\bf A}\tr{\bf A}$ (or ${\bf A}{\bf A}\tr$) contain the sums of squares of all the elements in the columns and in the rows, respectively, hence trace(${\bf A}\tr{\bf A}$) is the sum of squares of all the elements of {\bf A}.}
\vspace{-1cm}
\begin{flalign}
& \textrm{5.\ \  Sum-of-squared errors:} & \ & \hspace{-0.1cm}
 E  = {\rm trace}[({\bf F}_1^\ast-{\bf F}_2^\ast{\bf Q})\tr({\bf F}_1^\ast-{\bf F}_2^\ast{\bf Q})] &  \\[3pt]
& \textrm{6.\ \  Procrustes correlation:} & \ & \hspace{-0.1cm}
\mkern5mu r  = \sqrt{1 - E} & 
\end{flalign}

\noindent
The Procrustes correlation can be equivalently computed by vectorizing the two matrices ${\bf F}_1^\ast$ and ${\bf F}_2^\ast{\bf Q}$, and computing the Pearson correlation between them.

The function \texttt{protest()} \citep{Peres:01} in the R package \texttt{vegan} \citep{Oksanen:19} computes the correlation as follows:

\smallskip
\centerline{\texttt{protest(A, B, permutations=0)\$t0}}
\smallskip

\noindent
where \texttt{A} and \texttt{B} are the final matrices fitted to each other, in the above ${\bf F}_1^\ast$ and ${\bf F}_2^\ast{\bf Q}$.

\section*{Data and code availability}
\noindent
The Rabbit and original Crohn data are available on

\smallskip
\centerline{\tt https://github.com/michaelgreenacre/CODAinPractice}
\smallskip
\noindent
where the \textsf{R} code will also be available to reproduce most of the analyses. 

\bigskip

\bibliography{chiPower_arXiv}% common bib file



%%%%%%%%%%%%%%%%%%%%%%%%%%%%%%%%%%%%%%%%%%%%%%%%%%%%%%%%%%%%%%%%%%%%%%
%  SUPPLEMENTARY MATERIAL
%%%%%%%%%%%%%%%%%%%%%%%%%%%%%%%%%%%%%%%%%%%%%%%%%%%%%%%%%%%%%%%%%%%%%%
\newpage   % NEW

\centerline{\LARGE SUPPLEMENTARY MATERIAL}
\setcounter{table}{0}
\renewcommand{\thetable}{S\arabic{table}}

\setcounter{figure}{0}
\renewcommand{\thefigure}{S\arabic{figure}}

\setcounter{equation}{0}
\renewcommand{\theequation}{S\arabic{equation}}

%\newpage   %%% RESTORE WHEN BACK

\bigskip

\bigskip

%%%%%%%%%%%%%%%%%%%%%%%%%%%%%%%%%%%%%%%%%%%%%%%%%%%%%%%%%%%%%%%%%%
\noindent
\textbf{\large S1. Effect of zero replacement}

\medskip

\noindent
As remarked in Section 2.1, the data set \texttt{Crohn} exists in two different versions: the original version with zeros in the package \texttt{selbal}, used by \citep{Rivera:18}, and a modified version with no zeros in the more recent package \texttt{coda4microbiome} \citep{Calle:23}.  \cite{Rivera:18} use the original data and replace the zeros using function \texttt{cmultRepl()} in the package \texttt{zCompositions} \cite{Palarea:15}. The same zero replacement strategy was used by \cite{Coenders:22}.
Apart from the GitHub site mentioned in the under ``Data and code availability" in the present paper, the original version of the data is retrievable at this link:

\centerline{\tt https://github.com/UVic-omics/selbal}

\noindent
which explains the installation of the \texttt{selbal} package.
The package contains the original data object \texttt{Crohn}, with zeros, which has the bacterial predictors in the first 48 columns and the binary variable indicating Crohn's disease in column 49. 
The zero values are thus counted as follows:

\begin{verbatim}
  data(Crohn)
  sum(Crohn[,1:48]==0)
  # [1] 13474
\end{verbatim}

\noindent
The 13474 zero values represent 28.8\% of the data set. 

\begin{figure}[ht]
\begin{center}
\includegraphics[width=11.2cm]{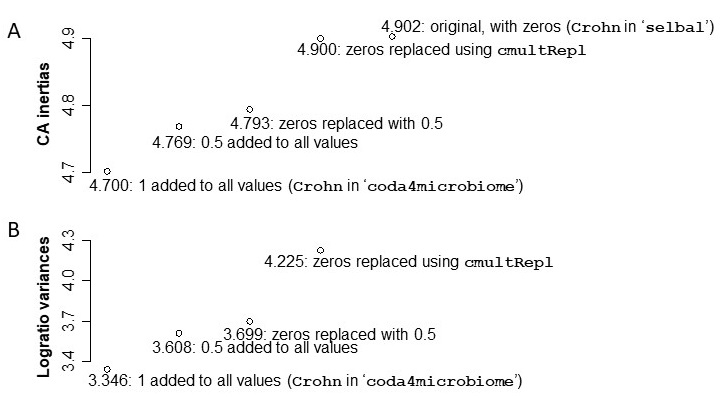}
\caption{Comparison of (A) inertias in correspondence analysis, and (B) logratio variances for different zero replacement strategies.}
\label{Supp_Inertias}
\end{center}
\end{figure}

By contrast, the modified version of these data, published by \cite{Calle:23} in the \texttt{coda4microbiome} package, with the same name \texttt{Crohn}, contains the modified version with the value of 1 added to every cell of the original $975\times 48$ data matrix. 
Here the effect of this method of zero avoidance is compared with some alternative strategies:
\begin{itemize}
    \item 
    adding 0.5 to the whole data matrix; 
   \item 
   substituting the zeros with 0.5;
   \item 
   the zero imputation approach using function \texttt{cmultRepl()}.
\end{itemize}
It is not possible to use the total logratio variance to compare these different strategies with the original \texttt{Crohn} data set, with zeros. 
Correspondence analysis (CA), however, which is close to logratio analysis, handles data zeros, and the total variance in CA, called the total inertia, can be computed for all versions of this compositional data set, with or without zero replacement.
This provides some measure of comparison of the effect of the different ways of modifying the data, which is done to allow logratio methods to be applied.
Figure \ref{Supp_Inertias}A shows the five inertia values in ascending order. 
The highest value is for the original data set, with zeros, which is expected, since any replacement of the zeros will reduce the variation in the compositional profiles.
But there is only a tiny reduction in inertia using \texttt{cmultRepl} replacement, while the others reduce the inertia more, with the modified data set published in \texttt{coda4microbiome}, where 1s have been added to the whole matrix, reducing the inertia the most.

In Figure \ref{Supp_Inertias}B the logratio variances are shown, and they are in the same order and showing a similar pattern.
The lowest is for the modified data set in \texttt{coda4microbiome}, and the highest is for the original version with zeros replaced using \texttt{cmultRepl}. 
Here it is impossible to compute a value for the logratio variance of the original data set, with zeros, but the results in Figure \ref{Supp_Inertias}A give some credence to using the value of 4.225 as a minimum reference for comparing the others.
It is then apparent that the zero avoidance strategy of adding 1 to each count of the original data set has (at least) reduced the logratio variance from 4.225 to 3.346, i.e., a variance loss of 20.8\%.
%It would be interesting to compare other zero replacement strategies \citep{Lubbe21} in the same way and to assess the effect on the results of analyses using these different versions of the data.

\begin{figure}[ht]
\begin{center}
\includegraphics[width=7.5cm]{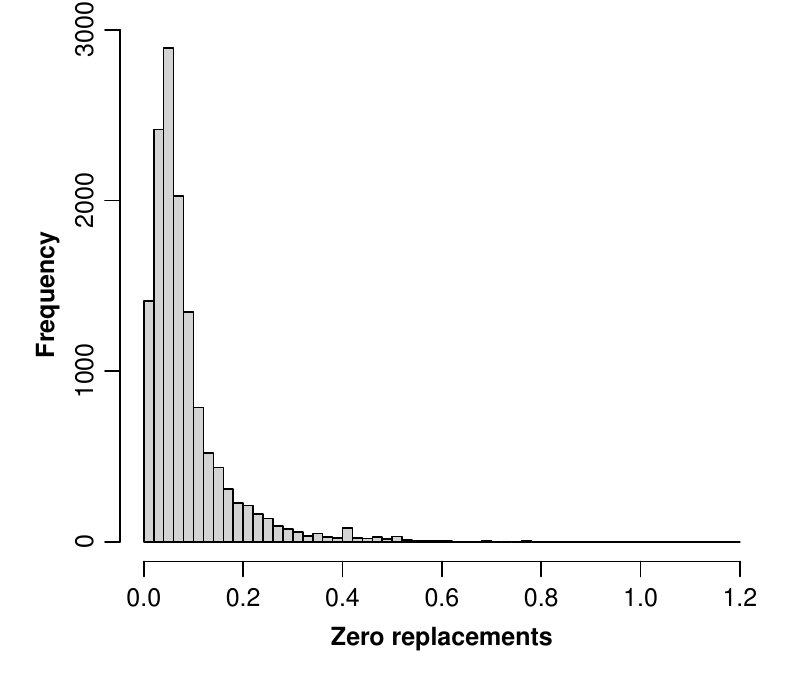}
\caption{Histogram of the zero replacements for the Crohn count matrix using function \texttt{cmultRepl}.}
\label{ZeroReplacements}
\end{center}
\end{figure}

The closeness of the data with zeros replaced using \texttt{cmultRepl} to the original data with zeros can be understood by examining the values of the 13474 replacements, shown as a histogram in Figure \ref{ZeroReplacements}.
99.4\% of the zeros have been replaced by values less than 0.5, which explains why the CA inertia is so close to the inertia of the original data.
Only one replaced value is greater than 1, equal to 1.15.

\vspace{-0.3cm}

\begin{figure}[H]
\begin{center}
\includegraphics[width=10cm]{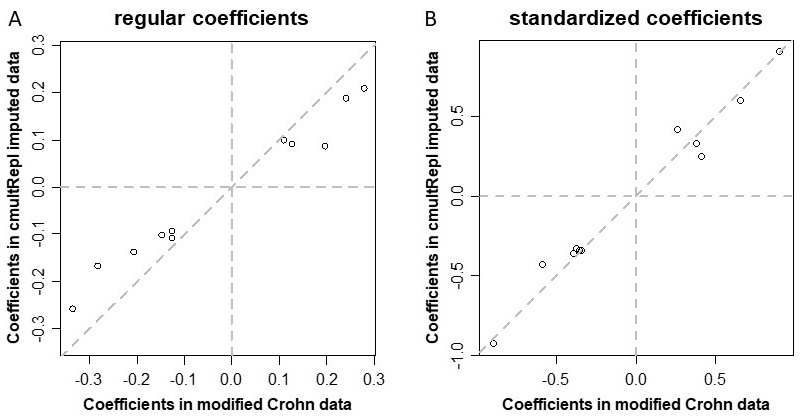}
\caption{Comparison of logistic regression coefficients for 11 logratio predictors, using the modified Crohn data (1 added to original counts), and respective coefficients when computed on the Crohn data with zeros replaced using function \texttt{cmultRepl}. A. Regular regression coefficients. B. Standardized regression coefficients}
\label{ModelsZeroRepl}
\end{center}
\end{figure}

\vspace{-0.3cm}

It would be interesting to compare other zero replacement strategies \citep{Lubbe21} in the same way and to assess the effect on the results of analyses using these different versions of the data.
As an example, consider the model in the first line of Table \ref{Predictions}, namely the logistic regression predicting Crohn's disease using 11 logratios computed on the modified Crohn data, with 1s added to the whole matrix. 
The logistic regression was repeated for the same 11 logratios computed this time on the data with zeros replaced using \texttt{cmultRepl}, as discussed in this section. 
The two sets of coefficients are compared in the scatterplot of Figure \ref{ModelsZeroRepl}, using first the regular regression coefficients and then the standardized ones. 
Comparing this result with Figure \ref{Coherence} it is apparent that the way zeros are replaced in the `coherent' logratio approach can produce as much change in the results as the formation of subcompositions in the `incoherent' chiPower approach.
An in-depth study of the effect of zero replacement on results using logratios would provide more insight into this issue.
.

\bigskip

\bigskip

%%%%%%%%%%%%%%%%%%%%%%%%%%%%%%%%%%%%%%%%%%%%%%%%%%%%%%%%%%%%%%%%%%
\noindent
\textbf{\large S2. Link between correspondence analysis and logratio analysis}

\medskip

\noindent
It is known that the principal component analysis (PCA) of all pairwise logratios, called logratio analysis (LRA) \citep{Greenacre:18}, and correspondence analysis (CA) of the compositional data matrix \citep{Greenacre:16a}
have a theoretical connection via the Box-Cox power transformation (\ref{BoxCox}), when data are all strictly positive.
This close connection between these two apparently different methods can be understood as follows -- see \cite{Greenacre:09, Greenacre:10a} for more detailed explanations.
Both methods are based on the singular-value decomposition (SVD) and both methods double-center the data matrix, so when it comes to Box-Cox transforming the data, the term $-\frac{1}{\lambda}$ can be omitted, so the transformation is simply $\frac{1}{\lambda} x^\lambda$.
In CA the matrix being decomposed is the weighted double-centering of the data matrix {\bf X} divided by its grand total $x_{++}$, where the weights are proportional to the row and column margins of the data matrix.
For a compositional data matrix the row weights are initially the same, whereas the column weights are different, but as the power $\lambda$ applied to the compositional data reduces towards zero, the column weights tend to uniformity, that is, the double-centering tends to being unweighted.
On the other hand, in LRA the matrix being decomposed is the unweighted double-centering of the log-transformed compositional data set $\log(\textbf{X})$.
Since the power-transformed matrix in CA is converging to $\log(\textbf{X} / x_{++}) = \log(\textbf{X}) - \log(x_{++})$ and the additive constant $-\log(x_{++})$ is eliminated by the double-centering, the limit is indeed the SVD of the double-centered $\log(\textbf{X})$, that is, LRA.

This result only applies to strictly positive data, although we will soon treat the case when zeros are present.
The multiplicative rescaling by $\frac{1}{\lambda}$ is important to include since it corrects for the decreasing variance as $\lambda$ decreases.  
This result can be re-expressed equivalently as the PCA of the chiPower-transformed data converging to LRA (see the next section, Section S2).
Notice that LRA has a weighted and unweighted version \citep{Greenacre:18} --  for the purposes of the present study the unweighted version is used, as in \cite{AitchisonGreenacre:02}.

The important implication of the above result is that, for strictly positive data, one can perform PCA with a very strong chiPower transformation (i.e., tiny power $\lambda$), and come as close as required to performing LRA based on all the LRs.
Equivalently, the Euclidean distances on chiPower-transformed compositional data (i.e., chi-square distances) converge to the logratio distances as the power tends to 0.
%In other words, the chiPower transformation converges to a transformation that is isometric with respect to the CLR transformation, which defines the logratio geometry.
Thus, in coherence terms, the chiPower transformation is increasingly coherent as the power decreases and tends to exact coherence (when all compositional data are positive).

\begin{figure}[ht]
\begin{center}
\includegraphics[width=9.5cm]{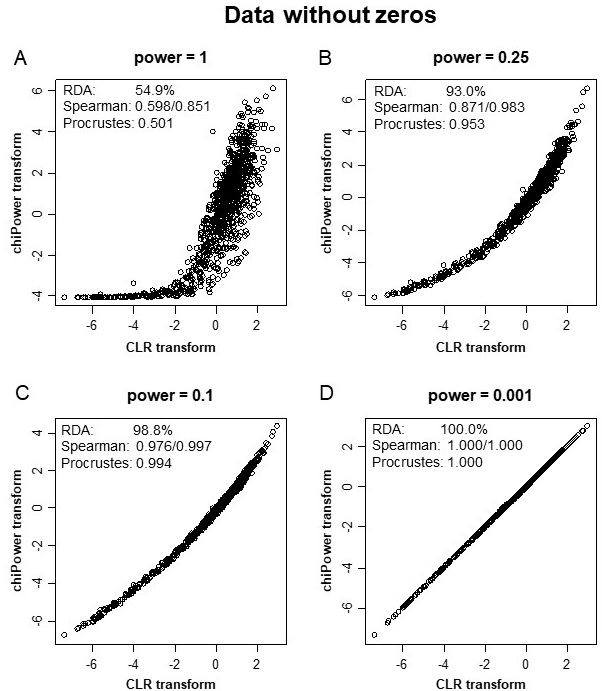}
\caption{Performance of chiPower transforms with powers of 1, 0.25 (fourth-root), 0.1 and 0.001, showing the approximation to the CLR transform for the eighth compositional part, \emph{ Bacteriodes}, of the modified Crohn data, without zeros. For power equal to 1 (subfigure A), this is equivalent to the chi-square standardization in CA that leads to chi-square distances. For the strong one-thousandth root power transformation (subfigure D), chiPower is indistinguishable from the CLR transformation, on all three statistics. The summary statistics (computed for the whole $975\times 48$ data matrix) are RDA: the percentage of logratio variance explained by the chiPower transform, as computed in an RDA; Spearman rank correlation between the two transforms; and the Procrustes correlation between the logratio geometry and the chiPower geometry. } 
\label{Crohn4transforms_nozeros}
\end{center}
\end{figure}

Figure \ref{Crohn4transforms_nozeros} shows the relationship between the chiPower transform, for four different powers, and the CLR transform, for the modified Crohn without zeros. The values plotted correspond to the 975 values for the eighth compositional part, {\it Bacteriodes}, of the Crohn data.
The summary statistics show how explained CLR variance, Spearman and Procrustes correlations change as the power is reduced to near 0.
The first Spearman correlation is between the logratio distances and the chiPower distances, and the second between the plotted values for the eighth compositional part.

\begin{figure}[ht]
\begin{center}
\includegraphics[width=9.5cm]{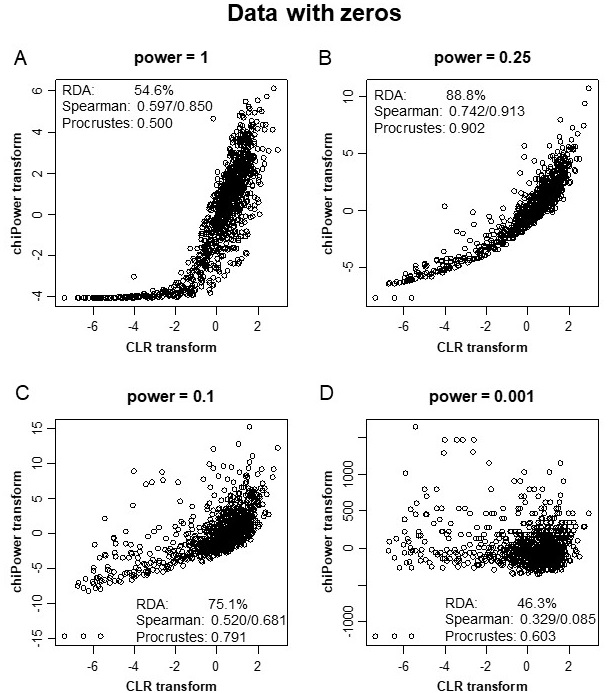}
\caption{Performance of chiPower transforms with powers of 1, 0.25 (fourth-root), 0.1 and 0.001, showing the approximation to the CLR transform, for the eighth compositional part, \emph{ Bacteriodes}, of the original Crohn data with zeros. For power equal to 1 (subfigure A), this is equivalent to the chi-square standardization in CA that leads to chi-square distances, for the original data, (cf. Figure \ref{Crohn4transforms_nozeros} for the modified data). The breakdown of the chiPower transform is apparent for smaller powers, since the approximation to the logratio transfomr is no longer possible. The summary statistics (computed for the whole $975\times 48$ data matrix) are RDA: the percentage of logratio variance explained by the chiPower transform, as computed in an RDA; Spearman rank correlation between the two transforms (where the second correlation is for the the bivariate scatterplot shown); and the Procrustes correlation between the logratio geometry and the chiPower geometry. } 
\label{Crohn4transforms_withzeros}
\end{center}
\end{figure}

Figure \ref{Crohn4transforms_withzeros} shows the relationship between CLR transform and the chiPower transform, now using the original Crohn data with zeros for the chiPower transform, for the same four different powers. As in similar comparisons, the CLR transform is necessarily computed using the modified Crohn data without zeros. The plotted values again correspond to the 975 values for the eighth compositional part, {\it Bacteriodes} (the few zero values can now be clearly seen in the subplots B, C and D).
Figure \ref{Crohn4transforms_withzeros}B is for the optimal power of 0.25 according to the criterion of maximizing the Procrustes correlation, with maximum 0.902.
The deterioration in the values of the statistics can be seen in subplots C and D.

\bigskip

\bigskip

%%%%%%%%%%%%%%%%%%%%%%%%%%%%%%%%%%%%%%%%%%%%%%%%%%%%%%%%%%%%%%%%%%
\noindent
\textbf{\large S3. Equivalence of CA and PCA for a compositional data matrix with chi-square standardization and optional powering}

\medskip

\noindent
Before introducing the power transformation, it is first shown that a CA of a compositional data matrix is equivalent to a PCA of the compositions with chi-square standardization.

Suppose $\bf X$ ($I\times J$) is a closed compositional data matrix, with all rows summing to 1, hence with a grand total equal to the number of rows $I$.
One definition of CA is that it is obtained from the following SVD \citep{Greenacre:16a}:
\begin{equation}
{\bf D}_r^{-\frac{1}{2}} \left( {\bf P} - {\bf r c}\tr \right) {\bf D}_r^{-\frac{1}{2}} = {\bf U D}_\phi {\bf V}\tr
\label{CA_SVD}
\end{equation}

\noindent
where:
\begin{itemize}
  \item   
  ${\bf P}$ is the data matrix divided by its grand total, in this case $(1/I)\bf X$ ;
  \item 
  {\bf r} and {\bf c} are the row and column marginal sums of $\bf P$; in this case ${\bf r} = (1/I)\bf 1$ and ${\bf c} = \bar{\bf x} = [\  \bar{x}_1 \ \bar{x}_2 \ \cdots \  \bar{x}_J \ ]\tr$, the vector of column means of $\bf X$ ;
  \item ${\bf D}_r$ and ${\bf D}_c$ are diagonal matrices of {\bf r} and {\bf c}; in this case ${\bf D}_r^{-\frac{1}{2}}$ reduces to multiplying by $I^{\frac{1}{2}}$ and ${\bf D}_c^{-\frac{1}{2}} = {\bf D}_{\bar{x}}^{-\frac{1}{2}}$ ;
  \end{itemize}
Hence, (\ref{CA_SVD}) can be written as
\begin{equation}
  I^{\frac{1}{2}} \left( {\bf X}/I - (1/I){\bf 1 \bar{x}}\tr \right)  {\bf D}_{\bar{x}}^{-\frac{1}{2}} = {\bf U D}_\phi {\bf V}\tr
\label{CA_SVD2}
\end{equation}
and then simplified as
\begin{equation}
  I^{-\frac{1}{2}} \left( {\bf X}{\bf D}_{\bar{x}}^{-\frac{1}{2}} - {\bf 1} \bar{\bf x}^{(\frac{1}{2}){\sf T}} \right) = {\bf U D}_\phi {\bf V}\tr
\label{CA_SVD3}
\end{equation}
where $\bar{\bf x}^{(\frac{1}{2})} = [\  \bar{x}_1^{\frac{1}{2}} \  \bar{x}_2^{\frac{1}{2}} \  \cdots \  \bar{x}_J^{\frac{1}{2}} \ ]\tr$ denotes the vector of square roots of the column means in $\bar{\bf x}$. 
The SVD in (\ref{CA_SVD3}) is exactly the PCA of the matrix of chi-square standardized compositions, ${\bf X}{\bf D}_{\bar{x}}^{-\frac{1}{2}}$, the row vectors of which have mean $\bar{\bf x}^{(\frac{1}{2}){\sf T}}$, where the scaling by $I^{-\frac{1}{2}}$ ensures the eigenvalues are variances, and equal to those of the CA in (\ref{CA_SVD})-- see \cite{GreenacreEtAl:23}.

If the compositions are initially chiPower-transformed and then closed, the above steps are the same except the Box-Cox rescaling by $\frac{1}{\lambda}$ (see (\ref{BoxCox})) needs to be re-introduced into the CA solution.
This is because CA eliminates any scale change in the original data matrix, and so the singular values in (\ref{CA_SVD}) will be decreasing with decreasing $\lambda$.
This scale re-adjustment is done by multiplying the CA singular values by $\frac{1}{\lambda}$. 
This additional rescaling step is avoided by performing PCA on the chiPowered compositions, which maintain the scalar multiple $\frac{1}{\lambda}$ throughout the process.

\bigskip

\bigskip

\newpage
%%%%%%%%%%%%%%%%%%%%%%%%%%%%%%%%%%%%%%%%%%%%%%%%%%%%%%%%%%%%%%%%%%
\noindent
\textbf{\large S4. Biplots of Crohn data}

\medskip

\begin{figure}[ht]
\begin{center}
\includegraphics[width=11.8cm]{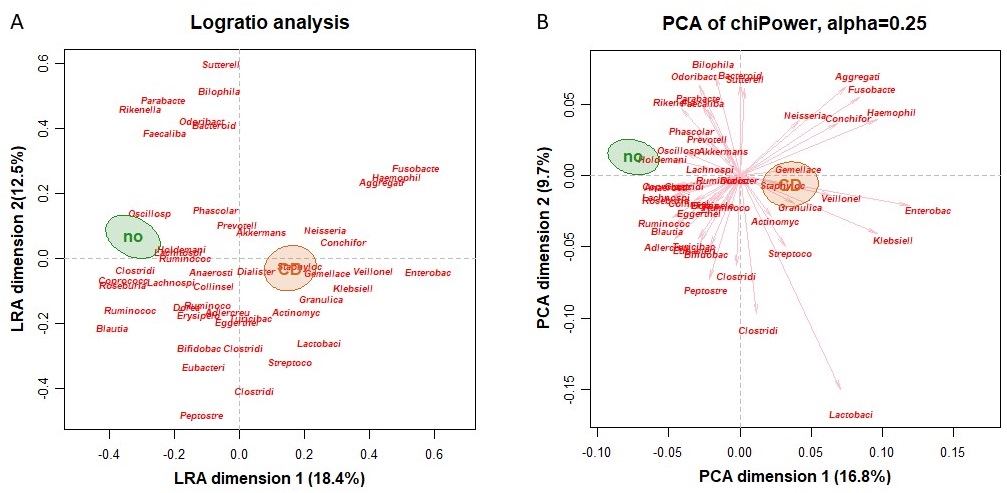}
\caption{Biplots of Crohn compositions, showing the 95\% confidence ellipses of the group means. A. Logratio analysis biplot (PCA of CLRs). B. PCA of chiPower transformed compositions with power 0.25. The sample configurations are not shown but their Procrustes correlation is 0.902 in the full space, increasing to 0.917 in the respective two-dimensional solutions shown here.} 
\label{CrohnLRACA}
\end{center}
\end{figure}

\noindent
Using the modified version of the Crohn data, where 1 has been added to the whole count matrix, the coordinates of the samples in 47-dimensional logratio space are computed.
Then, for $\lambda=0.01, 0.02, \ldots, 1$, the sample coordinates of the PCA of the chiPower-transformed data are computed and compared to the logratio coordinates using the Procrustes correlation.  
The maximum correlation of 0.902 is achieved with $\lambda=0.25$, i.e. a fourth-root.
Figure \ref{CrohnLRACA} shows the two solutions in their principal planes.

Be aware that the different configurations of the parts (bacteria) in Figure \ref{CrohnLRACA} do not reflect differences in the two treatments of the data, since the parts in the logratio analysis reflect pairwise logratios between pairs of parts, whereas in the PCA they reflect single variables, the power-transformed parts.
The assessment of the isometry is made on the sample geometries, not shown here since there are 975 samples, where the two confidence ellipses for the sample group means in each result are practically identical.

\bigskip

\bigskip
%%%%%%%%%%%%%%%%%%%%%%%%%%%%%%%%%%%%%%%%%%%%%%%%%%%%%%%%%%%%%%%%%%
\noindent
\textbf{\large S5. Coherence of Crohn data in unsupervised learning}

\medskip

\noindent
As explained in Section 2.1 and again in Supplementary Section S1, the Crohn data set exists in two versions: the original data set, which contains zeros, and the modified data set, which is the original data set to which 1 has been added to every value. To validate the chiPower transformation, the original data set is used, chiPower-transformed for the range of powers $0< \lambda \leq 1$, and its closeness to the logratio geometry of the modified data set is measured.
The result is an optimal power of $\lambda=0.25$, giving a Procrustes correlation of 0.902, as reported in the previous section..
The result turns out to be dependent on the zero replacement.
For example, if the zero counts are replaced using the function \texttt{cmultRepl} in package \texttt{zCompositions} \citep{Palarea:15}, which is a popular strategy in CoDA, the optimal power is $\lambda=0.18$, giving a higher Procrustes correlation of 0.948.
In Supplementary Material Section S1 it is shown that the zero replacement using \texttt{cmultRepl} is closer to the original data with zeros, which implies that the chiPower approach will give improved isometry, which is the case here. 

\begin{figure}[h]
\begin{center}
\includegraphics[width=9cm]{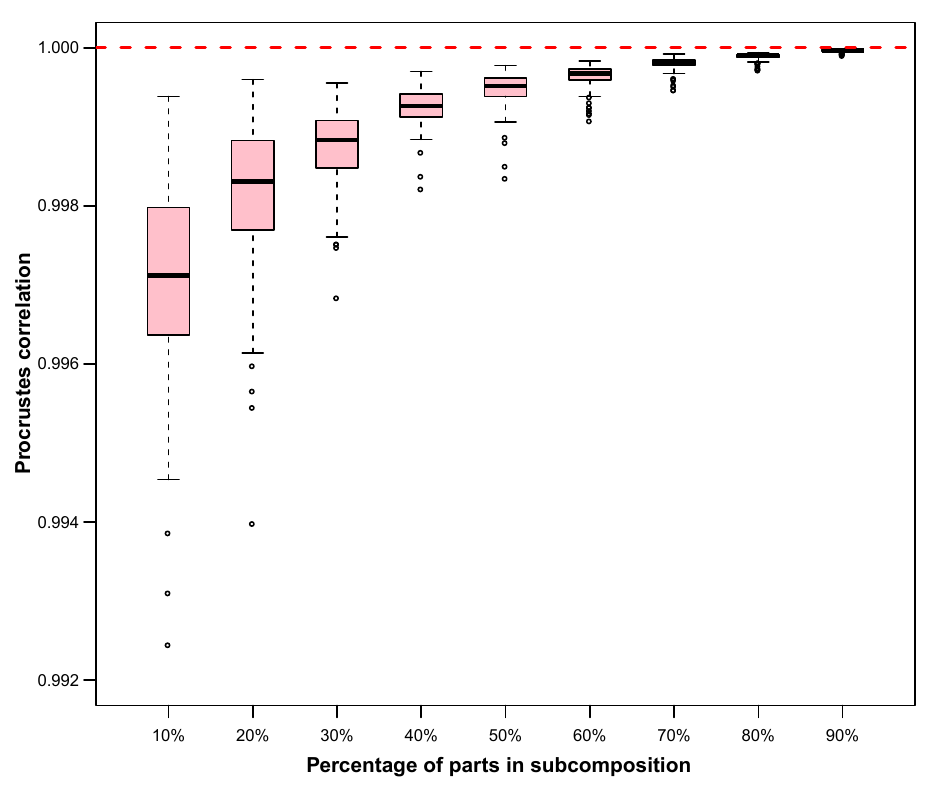}
\caption{Deviation from exact coherence of chiPower transformed Crohn data, with power $\lambda = 0.25$ (optimal for isometry). 1000 subcompositions are randomly generated, of sizes 10\% to 90\% of the 48-part Crohn data (original Crohn data, with zeros). For each size of subcomposition, a boxplot is shown of the Procrustes correlations with the same parts in the full composition.}
\label{Coherence_Crohn}
\end{center}
\end{figure}

Using the value $\lambda = 0.25$ the coherence is assessed using many random subcompostiions of different sizes, as explained in Section 2.5 and already applied to the Rabbits data in Section 3.2, 
The result is shown in Figure \ref{Coherence_Crohn}. 
The Procrustes correlations are still very high, but the deviation from coherence is larger than for the Rabbits data (Figure \ref{Coherence_Rabbits}B).
This is no doubt due to the fewer parts in the Crohn composition (48 here, compared to 3937 for the Rabbits data), which results in greater changes in the subcompositional values when closed.  

%%%%%%%%%%%%%%%%%%%%%%%%%%%%%%%%%%%%%%%%%%%%%%%%%%%%%%%%%%%%%%%%%%
\clearpage

%\medskip

%\noindent

\end{document}